\documentclass[preprint,showpacs,floats,floatfix,aps,12pt]{revtex4}
\usepackage{times} 
\usepackage{amssymb}
\usepackage{amsmath}
\ifx\pdfoutput\undefined
\usepackage{graphicx}     
\else
\usepackage{graphicx}
\fi
\usepackage{color}
\graphicspath{{.},{./fig/}} 
\newcommand{\mylab}[1]{\label{#1}}
\newcommand{\myfig}[1]{Fig.~\ref{fig:#1}}

\newcommand{\mysec}[1]{section~\ref{sec:#1}}
\usepackage{ulem}
\newcounter{countuwe}

\newcounter{countfil}

\begin{document}
%
\title{Rayleigh and depinning instabilities of forced liquid ridges on heterogeneous substrates}
\author{Philippe Beltrame}
\email{Philippe.Beltrame@univ-avignon.fr}
\affiliation{UMR EmmaH 1114, D\'epartement de Physique, Universit\'e
  d'Avignon, F-84000 Avignon, France}

\author{Edgar Knobloch}
\email{knobloch@berkeley.edu}
\affiliation{Department of Physics, University of California, Berkeley
  CA 94720, USA}

\author{Peter H\"anggi}
\email{peter.haenggi@physik.uni-augsburg.de}
\affiliation{Institut f\"ur Physik, Universit\"at Augsburg, D-86135 Augsburg,
Germany}
\author{Uwe Thiele}
\email{u.thiele@lboro.ac.uk}
\homepage{http://www.uwethiele.de}
\affiliation{Department of Mathematical Sciences, Loughborough University,
Loughborough, Leicestershire, LE11 3TU, UK}
\begin{abstract}
Depinning of two-dimensional liquid ridges and three-dimensional drops on
an inclined substrate is studied within the lubrication approximation.
The structures are pinned to wetting heterogeneities arising from variations
of the strength of the short-range contribution to the disjoining 
pressure. The case of a periodic array of hydrophobic stripes transverse to 
the slope is studied in detail using a combination of direct numerical 
simulation and branch-following techniques. Under appropriate conditions the 
ridges may either depin and slide downslope as the slope is increased, or first 
breakup into drops via a transverse instability, prior to depinning. The 
different transition scenarios are examined together with the stability 
properties of the different possible states of the system.  
\end{abstract}
%
%
\maketitle
%
%
%
\section{Introduction} \mylab{intro}
%
Experiments on sliding drops on solid substrates or on moving contact lines
show that the driving force must exceed a nonzero threshold in order that
motion results \cite{Duss79,deGe85,QAD98,BED07}. These observations are at 
variance with the theoretical predictions obtained for ideally smooth and 
homogeneous substrates for which motion starts for arbitrarily small driving 
force~\cite{deGe85,Thie01,Egge05,Egge05b,PiTh06,JoTh10}. This discrepancy 
between experiment and theory is believed to be caused by chemical 
heterogeneities and/or topographic roughness that are always present on
real substrates \cite{ScEl98,Jode84,NaGa94,Marm96,CDDR97}. A finite force
is then needed to depin the contact line or entire drop from the 
heterogeneity \cite{ScGa85,NaGa94,Marm96,JoRo90,QAD98,ScWo00,RoFo01}.

More generally, micro- or mesoscale heterogeneities are expected to affect the
macroscopic movement of drops. For instance, they are responsible for
contact angle hysteresis \cite{Duss79,deGe85,LeJo92}, the roughening
of contact lines \cite{Duss79,deGe85,RoJo87,ErKa94,GoRa01}, and the
stick-slip motion of weakly driven contact lines \cite{ScWo00,Tava06}.
Heterogeneities may also slow down or stop entirely the long-time coarsening 
of drop patterns in dewetting~\cite{Rehs01,BKTB02,TBBB03}.

These considerations apply to drops or liquid fronts (contact
lines) moving down an inclined substrate due to a gradient of
potential energy \cite{QAD98,PFL01,Thie01,LDL05}. However, they 
also apply to motion of drops or fronts caused by a
temperature gradient along the substrate (resulting in Marangoni
forces due to surface tension gradients) \cite{BBR93,Broc89}, or by
wettability gradients \cite{Raph88,ChWh92,ScEl98,DCC01,DaCh02,PiTh06}.

One approach to real (i.e., non-idealized) substrates is to consider
the limit of random heterogeneities
\cite{RoJo87,ErKa94,GoRa01,MGR04,LWRG06}.  Another approach focuses on
the effect of individual well-defined defects
\cite{Jode84,Rade89,MaCa93,MGR04}. Recently, the latter approach was
extended to study the depinning dynamics of drops on substrates with a
periodic array of precisely specified defects
\cite{ThKn06,ThKn06b,BHT09}. The approach employs a thin film
evolution equation with a spatially modulated disjoining pressure and
enables one to (i) study the depinning transition employing tools from
dynamical systems theory and bifurcation theory, and (ii) investigate
the dynamics of the stick-slip motion that occurs after depinning on
substrates with many defects. Contact line and drop motion on
regularly patterned substrates \cite{KoDi02,KoDi03,CuFe04,ZhMi09} is
in fact of considerable importance in various micro-fluidic and
nanotechnological applications
\cite{GCF00,Kim01,DDTR03,KJYB03,DCDT04}.  Single-cell assays, i.e.,
miniaturized devices for cell biology that consist of chemically
and/or physically structured substrates, provide a good example. In
this device each cell may be confined in an individual 'reaction
chamber', e.g., a drop of water on a hydrophilic spot. A pattern of
such spots allows for parallel analysis of a large number of
cells~\cite{LiAn10}. Driven drop motion on a regular heterogeneous 
substrate with a well-defined wettability period is also related to 
the motion of drops of partially wetting liquid on a horizontal 
rotating cylinder \cite{Thie10b}.

At first sight the qualitative behavior of driven drops on substrates with 
a well-defined array of defects described in previous studies of the problem
in a two-dimensional (2d) setting \cite{ThKn06,ThKn06b} agrees well with the 
results of three-dimensional (3d) computations \cite{BHT09}.  In a 2d system 
pinned drops can depin by two different mechanisms depending on the 
wettability properties of the defect, the drop size, and the driving force.  
The drops are either pinned by a hydrophilic defect at their back or by a 
hydrophobic defect at their front.  For the parameters investigated in
Ref.~\cite{ThKn06} the following depinning scenarios are found. In the 
hydrophilic case the pinned drop stretches quasistatically as the driving 
force increases but eventually loses stability through a 'sniper' (or 
Saddle-Node Infinite PERiod) bifurcation resulting in depinning. For forces 
larger than this critical force the drop slides in a periodic motion over the 
periodic array of defects. Theory implies that the mean drop speed beyond 
depinning should vanish as the square root of the distance from the sniper 
bifurcation \cite{Stro94} and this behavior is indeed observed in simulations. 
Each period of the resulting stick-slip motion consists of two distinct 
evolution phases that take place on distinct timescales: the drop slowly 
stretches away from the defect but once it breaks away it slides rapidly over 
to the next defect. The time scales for the 'sticking' and 'sliding' phases 
differ greatly close to the bifurcation, and the behavior that results 
resembles closely the experimentally observed stick-slip motion.  The 
situation is richer for hydrophobic defects that pin the drop by blocking it 
in front. In this case, in addition to the steady state sniper bifurcation, 
depinning can also occur via a Hopf bifurcation depending on the details of 
the defect \cite{ThKn06}. 

A recent study of the depinning of 3d drops from hydrophobic and hydrophilic 
line defects~\cite{BHT09} employing continuation and time-stepping 
algorithms~\cite{BeTh10} establishes a qualitative similarity between the 2d 
and 3d cases and supports the widely held expectation that studies of 2d thin 
film systems provide useful information about more realistic 3d systems.
However, significant differences do exist. These are mostly related to the 
additional degrees of freedom of the 3d system. For instance, the 3d drop can,
under appropriate circumstances, employ depinning pathways via morphological 
changes that are unavailable to a 2d drop. In the case of a hydrophilic defect 
the backward thread that connects the drop to the defect may gradually thin 
whereas in the hydrophobic case the drop may 'probe' the barrier locally by 
sending out an advancing protrusion over the defect~\cite{BHT09}. In fact, 
the 3d system has a number of features that have no counterpart in the 2d 
system, beyond the details of the depinning behavior of individual drops. 
The present study is dedicated to their analysis.  

Our starting point is the observation that all 2d drop solutions correspond 
in a 3d setting to spanwise-invariant ridge solutions. It follows therefore 
that \textit{all} results of~\cite{ThKn06,ThKn06b} remain valid in a 3d 
setting provided one imposes translation symmetry in the spanwise direction.
Thus stable ridge and drop solutions of identical liquid volume may coexist.  
Either of these solution types may become unstable as parameters are varied 
giving rise to branches of solutions that have not yet been studied. In the 
limit of zero driving (e.g., a horizontal substrate when gravity is the driving 
force) and a homogeneous substrate this question is related to the characteristics
of the Plateau-Rayleigh instability \cite{Plat1873,Rayl1879} of a static ridge 
(sometimes called a truncated cylinder) \cite{Davi80,Grin94}. 

In a driven 3d system on a heterogeneous substrate the question
translates into (i) an investigation of the lateral stability of
pinned and sliding spanwise-invariant ridges, and (ii) the
longitudinal stability of streamwise rivulets whose diameter is
modulated by the heterogeneities.  In both cases we expect changes of
stability that give rise to 'new' branches of (i) spanwise modulated
ridges and (ii) 'wavy rivulets', respectively. These states have no
counterpart in the 2d case.  We are interested in particular in the
relation between the stability of the rivulets \cite{WeDa81} and the
existence of depinned sliding drop solutions. The combination of these
aspects of the problem together with 2d and 3d depinning
characteristics obtained previously \cite{ThKn06,ThKn06b,BHT09}
provides a fairly complete picture of the 3d problem and of the
relation among static and sliding ridges, static and sliding
drops and steady or wavy rivulets.

The paper is organized as follows. In section~\ref{sec:mod} we
summarize the thin film model we use for the study of drop depinning
in 3d and introduce the numerical tools we employ. In
section~\ref{sec:res} we discuss the results, first for drops and
ridges on a horizontal substrate (section~\ref{sec:hori}), and then for
ridges on an inclined substrate (section~\ref{sec:incl-ridge}), followed by
drops and drop-like states on an inclined substrate 
(section~\ref{sec:dropridge}). We interpret our results in section 
\ref{sec:int}, followed by conclusions in section \ref{sec:conc}.

\section{Model and numerical tools}
\mylab{sec:mod}
%
\begin{figure} 
\includegraphics[width=0.9\hsize]{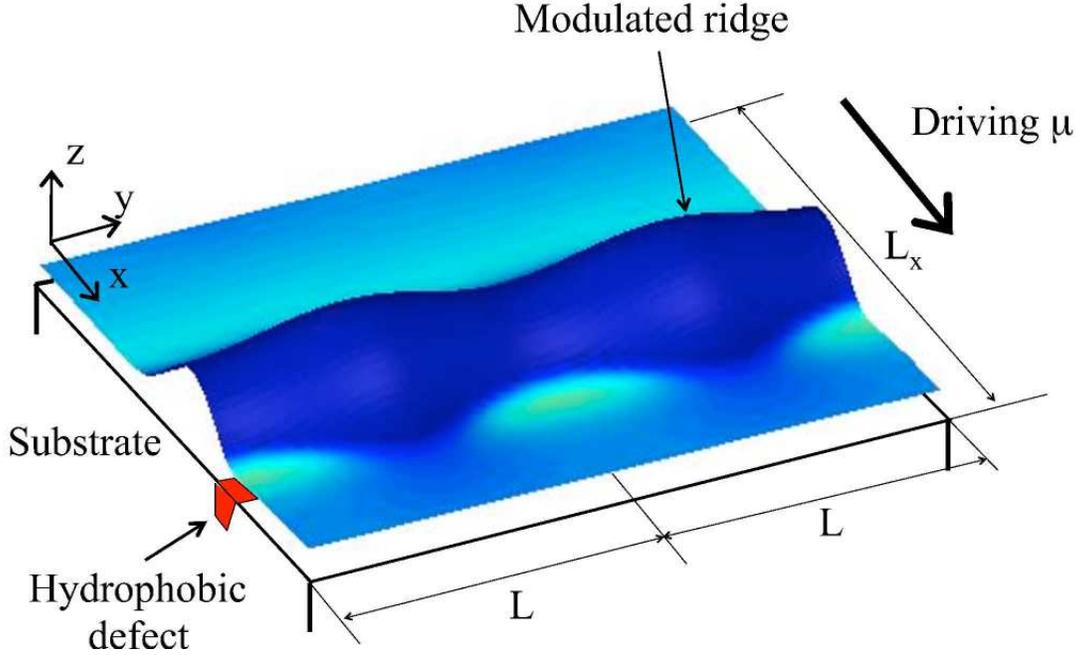}
\caption{(color online) Sketch of the geometry of the problem: A ridge or drop is
  pinned at a stripe-like hydrophobic defect, i.e., a heterogeneous
  wettability that depends only on the $x$-coordinate. A
  driving force $\mu$ acts in the $x$-direction.
  \mylab{fig:sketch}}
\end{figure} 

\subsection{Lubrication equation} 
The partial differential equation governing the time evolution of the
profile of a thin liquid film was discussed in depth in the 2d case
in Ref.~\cite{ThKn06} and adapted to the 3d case in Ref.~\cite{BHT09}. 
Briefly, we consider a layer or drop of partially wetting liquid (with 
a small equilibrium contact angle) on a flat chemically inhomogeneous
two-dimensional solid substrate (see sketch in Fig.~\ref{fig:sketch}).
Long-wave theory allows us to derive an evolution equation for the film
thickness profile $h(x,y,t)$ directly from the Navier-Stokes and
continuity equations \cite{ODB97,KaTh07}. We use no-slip and no penetration 
boundary conditions at the substrate, and the equilibrium of tangential and 
normal stresses at the free surface. The wettability properties are 
incorporated as a disjoining pressure~\cite{ODB97,deGe85}. In the presence 
of a small (for consistency with the long-wave approximation) lateral body
force in the $x$-direction we obtain the non-dimensional equation
\begin{equation} 
  \partial_{t}h=-\nabla\cdot\left\{ Q(h)\left[ 
\nabla\left(\Delta h+\Pi(h,x)\right)+\mu\mathbf{e_{x}} 
\right]\right\},
\mylab{eq:lub}
\end{equation} 
where $\nabla=(\partial_{x},\partial_{y})$ and 
$\Delta=\partial_{xx}^{2}+\partial_{yy}^{2}$ are the planar gradient
and Laplace operator, respectively, with $(x,y)$ denoting the downstream
and spanwise directions. The mobility function $Q(h)\equiv h^{3}$
corresponds to a parabolic velocity profile (Poiseuille
flow). Capillarity is represented by $\Delta h$ (Laplace
pressure). The position-dependent wetting properties are incorporated
via a $y$-independent disjoining pressure $\Pi(h,x)$ in order to focus
on stripe-like defects. Of the different functional forms for $\Pi(h)$ 
found in the literature \cite{deGe85,TDS88}, many allow for the presence 
of a precursor film of thickness 1-10 nm on a 'dry' substrate and these
are used to describe partial wetting. In this way 'true' film rupture
in dewetting and the stress singularity at a moving contact line are
avoided.  Here, we employ long-range apolar van der Waals interactions
combined with a short-range polar electrostatic or entropic interaction 
\cite{deGe85,Shar93,TVN01}, leading to the dimensionless disjoining pressure
\begin{equation} 
  \Pi(h,x)=\frac{b}{h^{3}}-\left[1+\epsilon\xi(x)\right]e^{-h},
\mylab{eq:dispin}
\end{equation} 
where the parameter $b>0$ indicates the relative importance of the two 
antagonistic interactions. In the following we employ this particular form
of the disjoining pressure with the parameter $b$ fixed at $b=0.1$
to allow for quantitative comparison with the results of 
Refs.~\cite{ThKn06,ThKn06b,BHT09} but emphasize that any 
qualitatively similar disjoining pressure $\Pi$ yields like results, as
becomes clear when comparing, for example, the dewetting results 
in \cite{TNPV02,TVN01,ShKh98,ShKh99,TVNP01,Beck03}.  In
Eq.~(\ref{eq:dispin}), $\epsilon$ and $\xi(x)$ are the strength of the
wettability contrast and the shape function of the heterogeneity,
respectively. For the latter we use a Jacobi elliptic function to model 
periodic arrays of localized defects, specifically
\begin{equation}
\xi(x)\,=\,2\{\mbox{cn}[2K(k)x/L_{x},k]\}^2 - \bar{C},
\mylab{eq:kappa3}
\end{equation}
where $k$ is the modulus of the elliptic function and $K(k)$ is the
complete elliptic integral of the first kind.  As $k\rightarrow 1$
Eq.~(\ref{eq:kappa3}) describes localized defects. Throughout the paper 
we take $k=(1 - 10^{-6})^{1/2}$ with $\bar{C}=0.241136$ so that
$\int_0^{L_x}\xi(x)dx=0$, i.e., so that the mean of the disjoining
pressure does not depend on the wettability contrast $\epsilon$. In
addition, we take the period $L_x$ sufficiently large to avoid 
interactions between adjacent drops/defects. The resulting wettability
profile $\xi(x)$ is shown in \myfig{ridge-diag-prof-hori}(b), lower
panel. The wettability contrast $\epsilon > 0$ [$\epsilon< 0$]
indicates a hydrophobic [hydrophilic] defect, i.e., it tells us
whether the defect is less [more] wettable than the surrounding
substrate. Thus $\Pi(h,x)$ allows us to incorporate a stripe-like wettability
pattern in the theory and to study the influence of chemical substrate
heterogeneities or defects via a spatial modulation of the material
parameters involved. This variation must take place on length scales
much larger than the film thickness for consistency with the long-wave
approximation \cite{Thie03}.

In nondimensionalizing the model to arrive at Eq.~(\ref{eq:lub}) we have used 
the length scales $l$ for the film thickness and $\sqrt{l\gamma/\kappa}$ for
the $(x,y)$ coordinates, the time scale $3\eta \gamma/\kappa^2 l$ for the
time and the force scale $\sqrt{l\gamma}/\kappa^{3/2}$ for the force.
The length $l$ corresponds to a characteristic scale for
the thickness of the wetting layer, while $\gamma$ and $\eta$ are the
surface tension and viscosity of the liquid, respectively; $\kappa$ is
a typical energy density scale related to wettability. The ratio of the 
vertical and horizontal length scales used, $\sqrt{l\kappa / \gamma}$, 
corresponds to the smallness parameter in the
lubrication approximation. This parameter is also closely related to
the equilibrium contact angle on a homogeneous substrate,
cf.\,\cite{PiTh06}. In writing Eq.~(\ref{eq:lub}) we assumed that the
lateral driving force, of dimensionless strength $\mu$, does not
depend on the film thickness. This is the case for gravitational or 
centrifugal forces. For driving forces that depend on the film thickness 
the mobility for the force will differ from that for the pressure term. 
This is the case for a purely lateral temperature gradient or a gradient 
of an electric field for a dielectric liquid in a capacitor.  Note that 
the long-wave scaling used here implies that the dimensionless contact 
angle and driving force $\mu$ may be of order one.

In the literature one finds two different ways of counting spatial
dimensions in the problem at hand. On the one hand, focusing on the
mathematical structure of Eq.~(\ref{eq:lub}) one distinguishes between 
one-dimensional ($h$ depends on $x$ only) and two-dimensional ($h$
depends on $x$ and $y$) cases. On the other hand, one may count
the physical dimensions and refer to the situation where the film
thickness depends only on $x$ [depends on $(x,y)$] as the two-dimensional
(2d) case [three-dimensional (3d) case]. Here we follow the
latter convention.

\subsection{Numerical schemes and parameters}
\mylab{sec:numpar}
Based on Eq.~(\ref{eq:lub}) and the disjoining pressure
(\ref{eq:dispin}) the 3d depinning behavior is analysed as follows:
Steady-state solutions (pinned drops) and their stability are
determined using continuation techniques and the stick-slip motion
beyond the depinning threshold is analysed using time-stepping
algorithms. Note that in the 2d case \cite{ThKn06,ThKn06b} an explicit
scheme suffices for the latter and continuation can be performed using
the package AUTO \cite{DKK91,DKK91b,AUTO2000}.  This is not possible
in the 3d case where an accurate and effective time simulation of
Eq.~(\ref{eq:lub}) remains a challenge, leading to a number of
numerical issues \cite{BGW01,Thie02,BeGr05}. For these reasons we employ 
here an exponential propagation approach based on the exact solution of 
the linearized equation at each time-step \cite{FTDR89}. The required
exponentiation of the Jacobian matrix is performed using a Krylov
reduction based on the Cayley-Arnoldi algorithm \cite{BeTh10}.  The
approach allows for a very good estimate of the optimal timestep in
both the slow and fast dynamical regimes. This is of paramount
importance since close to the depinning transition typical time
scales vary over many orders of magnitude.  The same Cayley-Arnoldi
algorithm is employed in our tangent predictor--secant corrector
scheme for the continuation of steady 3d drop states \cite{BeTh10}.
This approach is advantageous as it allows us to perform all tasks
arising in a bifurcation analysis simultaneously. This includes the
computation of the kernel of the Jacobian to detect bifurcations and
the stability analysis of the steady states.

Steady solutions are characterized by their $L^2$ norm 
\begin{equation}
||\delta h|| = \left[\frac{1}{L_x L_y}\int_0^{L_y}\int_0^{L_x}(h(x,y)-\bar{h})^2\,dx dy\right]^{1/2},
\mylab{eq:norm}
\end{equation}
while time-dependent states such as sliding drops or ridges are characterized 
by their temporal period $T$ and time-averaged norm
\begin{equation}
||\delta h|| = \left[\frac{1}{T}\,\frac{1}{L_x L_y}\,
\int_0^T\int_0^{L_y}\int_0^{L_x}(h(x,y,t)-\bar{h})^2\,dx dy dt\right]^{1/2}.
\mylab{eq:tnorm}
\end{equation}

In this paper we focus on a regular array of {\it hydrophobic} stripe-like 
defects with a wettability contrast $\epsilon = 0.3$ with periodic boundary 
conditions in the downstream direction with period $L_x=40$ containing a single
defect. Periodic boundary conditions are employed in the spanwise direction as 
well, with period $L_y$ that may be varied to study drops of different volume 
and spanwise ridges of different length. In these circumstances $L_y$ is used
as a bifurcation parameter and is then denoted by the symbol $L$. In this 
paper the values of $L$ are restricted by the requirement that the domain
contains no more than one drop in the spanwise direction. The mean film 
height is fixed at $\bar{h}=1.2$, compared with the precursor film height 
$h=1$. 
%
\section{Results}
\mylab{sec:res}

\subsection{Drops and ridges on a horizontal substrate: $\mu=0$} 
\mylab{sec:hori}
%
On a heterogeneous substrate with a periodic array of hydrophobic
defects the unique stable solution in the 2d case corresponds to a
2d drop sitting in the middle between two hydrophobic defects
\cite{ThKn06,ThKn06b,BHT09}. In the 3d case this solution corresponds
to a ridge with translational invariance in the spanwise 
direction. Such a ridge may, however, be unstable to a Plateau-Rayleigh
instability if the spanwise system size exceeds a critical value 
$L_c$ as occurs for ridges on homogenous substrates
\cite{Davi80,Grin94}.  For cylindrical liquid bridges between two
solids the primary bifurcation is an imperfect subcritical pitchfork
whose details depend on the particular setting (with/without gravity,
thermocapillarity etc.), the mode number, the boundary conditions at
the two supports and their geometry \cite{LoSt97,CRHS99}. Less
is known for a ridge on a solid substrate, where most results
\cite{Davi80,BKTB02,TBBB03,BKL05,KMW06,MRD08b} concern linear
stability. For example, ridges on striped heterogeneous substrates
can be stabilized w.r.t.\ the Plateau-Rayleigh instability for
any $L_y$ if they sit on a hydrophilic stripe or between hydrophobic
stripes of sufficient wettability contrast \cite{TBBB03}. In the
nonlinear regime we expect to find a (subcritical) pitchfork of
revolution (due to translation invariance in the spanwise
direction in our periodic setting) when the spanwise system size
$L\equiv L_y$ increases. Bulge states with large contact angles are
studied in \cite{BrLi02}.
\begin{figure}[tbh]
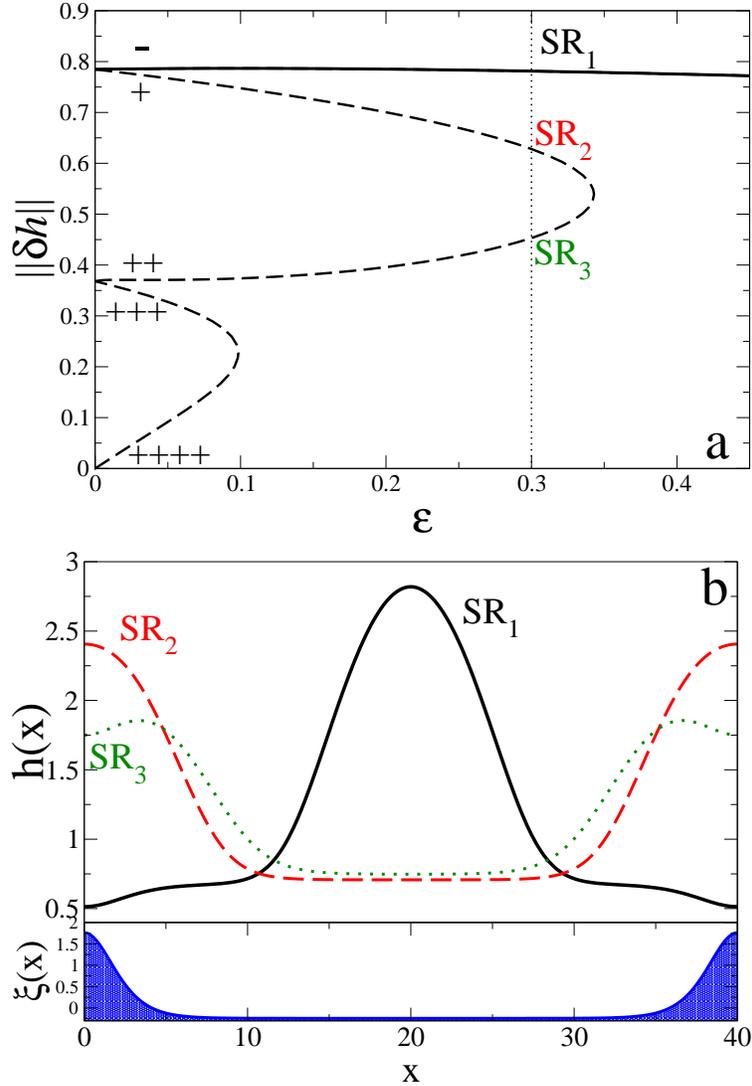

\includegraphics[width=0.6\hsize]{fig2a}\\[1ex]
\includegraphics[width=0.6\hsize]{fig2b}
\caption{(color online) (a) Steady ridge (SR) solutions on a horizontal substrate
  with hydrophobic line defects as a function of the wettability
  contrast $\epsilon$. Solid [dashed] lines indicate stable [unstable]
  solutions. The symbols $\pm$ indicate the stability of the branches 
  w.r.t.\ 2d perturbations, with $-$ indicating stability and the $+$'s 
  indicating the number of unstable eigenmodes.
  (b) Profiles of the solutions SR$_1$, SR$_2$ and SR$_3$
  at $\epsilon=0.3$ (top panel) together with the heterogeneity
  profile $\xi(x)$ (lower panel).  Here and in subsequent figures the
  hydrophobic defect is always centered at $x=0$ ($x=L$). Parameters: 
  ${\bar h}=1.2$, $L_x=40$.  }
\mylab{fig:ridge-diag-prof-hori}
\end{figure}

\begin{figure}[tbh]
\centering
\includegraphics[width=0.7\hsize]{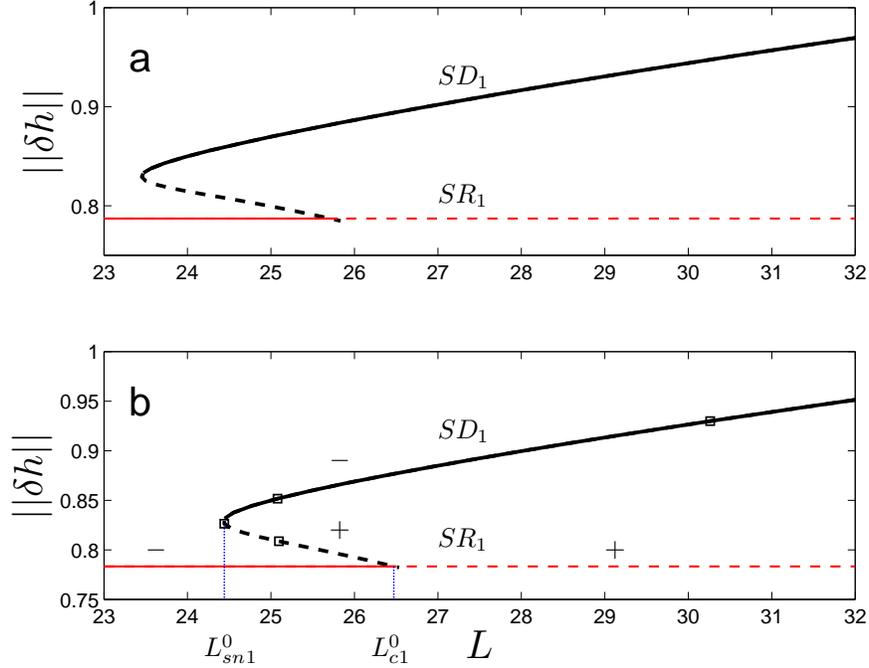}
\caption{(color online) Bifurcation diagrams for (a) $\epsilon=0$ and (b) $\epsilon=0.3$ showing 
the loss of stability of the SR$_1$ solutions (horizontal red line) with respect 
to 3d perturbations when the spanwise period $L$ increases together with 
the resulting branch of 3d states labeled SD$_1$ (black line). Solid [dashed]
lines indicate stable [unstable] solutions. Solution profiles at locations
indicated by open squares in panel (b) are shown in Fig.~\ref{fig:ridge-prof-hori}. 
Parameters: $\mu=0$, ${\bar h}=1.2$, $L_x=40$.}
\mylab{fig:celnob}
\end{figure}

\begin{figure}[tbh]
 \includegraphics[width=0.9\hsize]{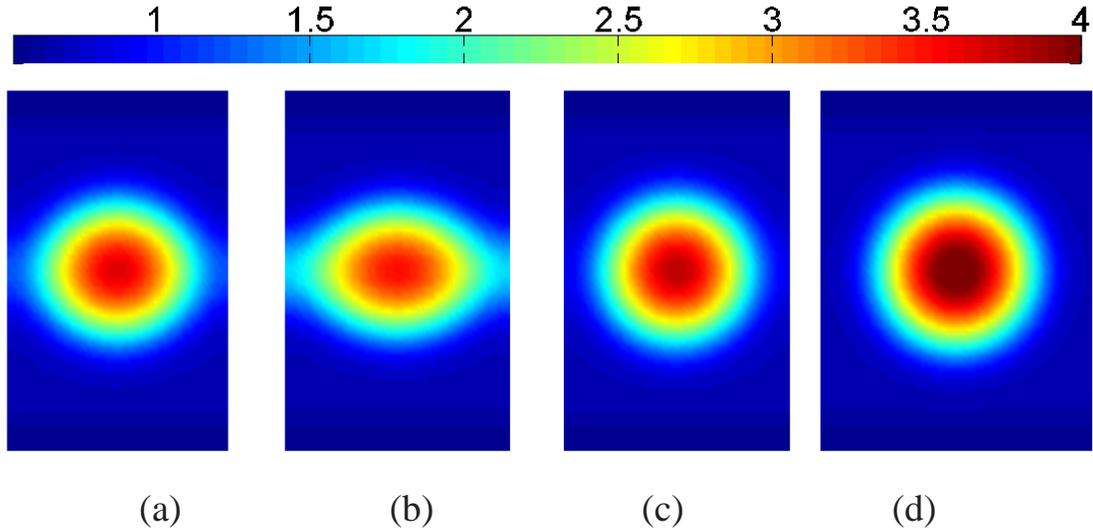}\\
{\large \hspace*{.1cm}\hfill (a)\hfill(b)\hfill(c)\hfill(d) \hfill\hspace*{.1cm}}
 \caption{(color online) Steady drop-like states SD$_1$ at locations indicated by open squares
in Fig.~\ref{fig:celnob}(b) in terms of contours of constant $h(x,y)$. (a) The 
saddle-node bifurcation at $L_\mathrm{sn1}^0= 24.44$, (b) the unstable branch 
at $L=25.09$, (c) the stable branch at $L=25.08$ and (d) $L=30.26$. 
The downslope direction $x$ is from top to bottom, with $y$ horizontal. 
Parameters: $\mu=0$, $\epsilon=0.3$, ${\bar h}=1.2$, $L_x=40$.}
\mylab{fig:ridge-prof-hori}
\end{figure}

\begin{figure}[tbh]
\centering
\includegraphics[width=0.7\hsize]{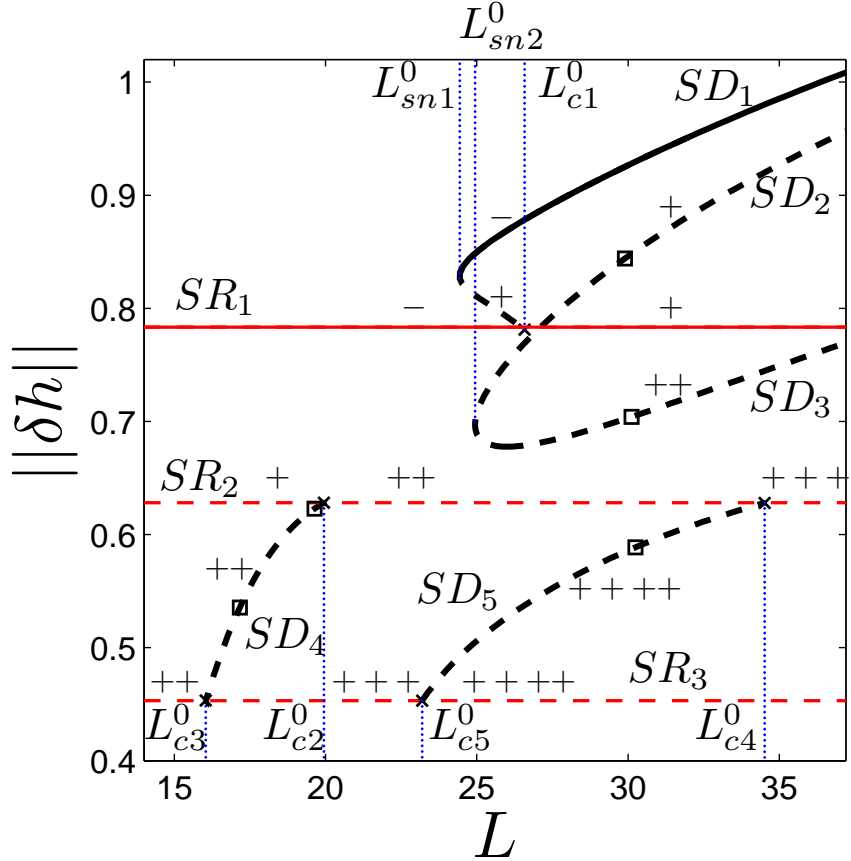}
\caption%
{(color online) Bifurcation diagram showing secondary bifurcations to 3d states from the branches 
SR$_1$, SR$_2$ and SR$_3$ (horizontal red lines) of 2d steady ridges, and the 
associated branches of 3d steady drop-like states SD$_1$, SD$_2$, SD$_3$ and SD$_4$ 
(heavy black lines). Solid [dashed] lines indicate stable [unstable] solutions. 
Solution profiles at locations indicated by open squares are shown in 
Fig.~\ref{fig:ridge-prof-hori2,3}. Parameters: $\mu=0$, $\epsilon=0.3$, 
${\bar h}=1.2$, $L_x=40$.}
\mylab{fig:celmu0eps0,3}
\end{figure}

\begin{figure}[tbh]
 \includegraphics[width=0.9\hsize]{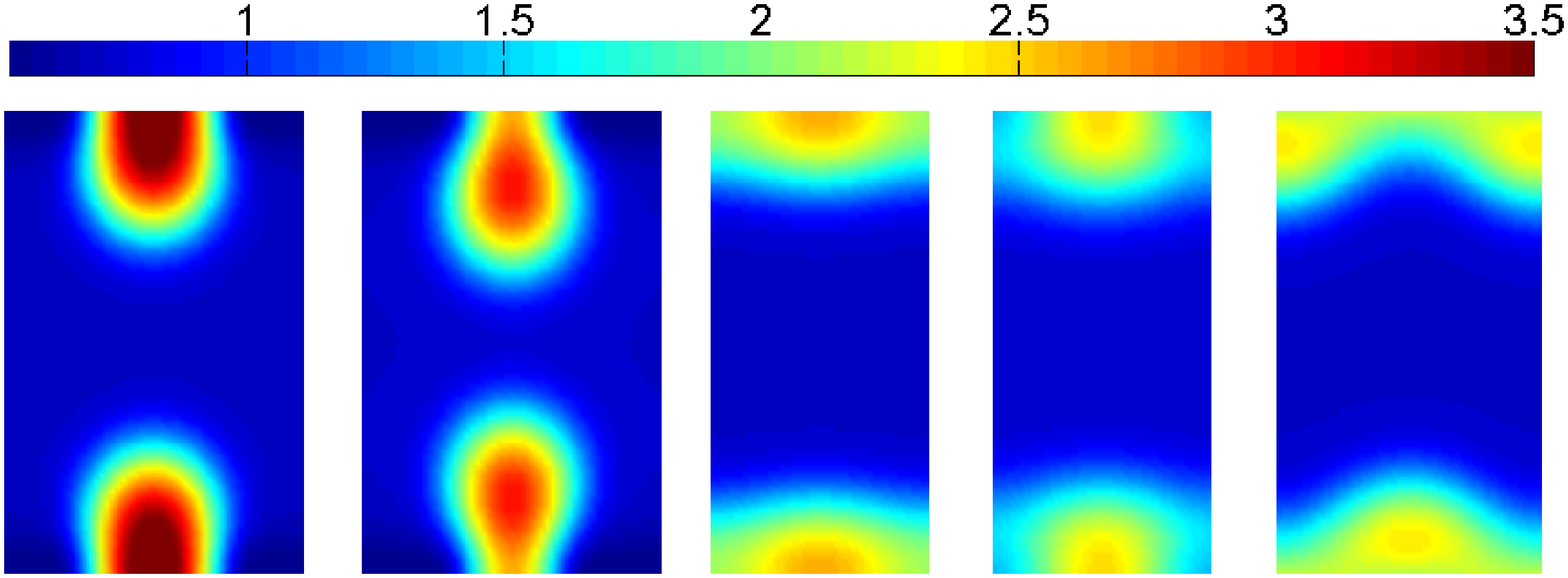}\\
{\large \hspace*{.1cm}\hfill (a)\hfill(b)\hfill(c)\hfill(d)\hfill(e) \hfill\hspace*{.1cm}}
\caption{(color online) Steady drop-like states at locations indicated
  by open squares in Fig.~\ref{fig:celmu0eps0,3} in terms of contours
  of constant $h(x,y)$.  (a) SD$_2$ at $L=29.90$, (b) SD$_3$ at
    $L=30.11$, (c) SD$_4$ at $L=19.64$ (close to SR$_2$), (d) SD$_4$
    at $L=17.17$ (close to SR$_3$), and (e) SD$_5$ at $L=30.25$. The
  downslope direction $x$ is from top to bottom, with $y$
  horizontal. Parameters: $\mu=0$, $\epsilon=0.3$, ${\bar h}=1.2$,
  $L_x=40$.}
\mylab{fig:ridge-prof-hori2,3}
\end{figure}

Figure \ref{fig:ridge-diag-prof-hori}(a) shows the 2d solutions on a horizontal
substrate ($\mu=0$) as a function of the wettability contrast $\epsilon$ and
indicates the presence of three 2d states when $\epsilon=0.3$. The solutions
take the form of steady ridges (SR) of different $L^2$ norm and differ in both
the location of the ridge relative to the hydrophobic heterogeneity and their 
linear stability properties. The largest amplitude solution, labeled SR$_1$, 
consists of a ridge confined midway between adjacent hydrophobic defects 
[Fig.~\ref{fig:ridge-diag-prof-hori}(b)]; this solution is stable with respect 
to 2d perturbations. In contrast, the solutions SR$_2$ and SR$_3$ are unstable. Of 
these, SR$_2$ consists of ridges superposed on top of the defects, a configuration 
that is expected to be unstable. The lowest amplitude solution SR$_3$ is 
characterized by minima in the film profile $h(x)$ at both the defects and half
way between them [Fig.~\ref{fig:ridge-diag-prof-hori}(b)]. This solution is 
also unstable.

Figure~\ref{fig:celnob} extends these results to 3d structures on a
horizontal substrate ($\mu=0$). To obtain the figure we start with a
stable SR$_1$ solution in Fig.~\ref{fig:ridge-diag-prof-hori} with
either $\epsilon=0$ (homogeneous substrate) or $\epsilon=0.3$
(hydrophobic line defect) and study its linear stability with respect
to 3d volume-conserving perturbations with period $L$ in the spanwise
direction. A symmetry-breaking Plateau-Rayleigh-like instability sets
in for $L>L_{c1}^0$ (the superscript zero refers to $\mu=0$, i.e., the
absence of a driving force). At $L_{c1}^0$ a branch of steady 3d
solutions bifurcates from the 2d states when an eigenvalue of double
multiplicity becomes unstable. The double multiplicity is a consequence
of O(2) symmetry of the SR stability problem under translations in $y$
modulo the period $L$ together with the reflection $y\rightarrow -y$ and
is not indicated in Fig.~\ref{fig:celnob} or subsequent figures. The
3d solutions that result take the form of unstable ridges modulated in
the spanwise direction [see Fig.~\ref{fig:ridge-prof-hori}(b)] and are
unstable. With increasing modulation amplitude these solutions turn
around at a saddle-node bifurcation (at $L=L_\mathrm{sn1}^0$) and acquire
stability. At approximately the same location the solution ceases to
resemble a spatially modulated ridge and begins to resemble a steady
drop-like (SD) state. In the following we use the notation SD to refer
to all 3d states, even near the bifurcation $L_{c1}^0$, where their
appearance is ridge-like.  It follows that for $L<L_\mathrm{sn1}^0$
the only stable solution is the 2d ridge state SR$_1$ that is invariant
w.r.t.\ spanwise translations. For $L>L_{c1}^0$ the ridge is linearly
unstable and decays into the drops SD$_1$ that constitute the only
stable solution in this regime. In between, i.e., in the range
$L_\mathrm{sn1}^0<L<L_{c1}^0$, both the SR$_1$ and the larger
amplitude SD$_1$ solutions are linearly stable, while the unstable
subcritical SD$_1$ branch of modulated ridges corresponds to unstable
threshold solutions separating the two stable solutions. Comparison of
Figs.~\ref{fig:celnob}(a,b) shows that presence of the defect shifts
the bifurcation to 3d states and the saddle-node bifurcation on the
resulting branch of 3d states to larger values of the parameter $L$
than required for the Plateau-Rayleigh instability on a homogeneous
substrate. In fact, this stabilizing effect is much more pronounced in the
hydrophilic case ($\epsilon<0$, not shown).

Figure~\ref{fig:ridge-prof-hori} shows a sequence of solutions along the SD$_1$ 
branch of 3d states created at $L_{c1}^0$ when $\mu=0$,
corresponding to locations indicated by open squares in Fig.~\ref{fig:celnob}(b),
i.e., for $\epsilon=0.3$. Figure~\ref{fig:ridge-prof-hori}(a) shows the solution 
at the saddle-node ($L_\mathrm{sn1}^0= 24.44$), with (b) showing the 
unstable ridge-like state below the saddle-node ($L=25<L_{c1}^0= 26.6$) 
and (c) showing the corresponding stable drop-like state above the saddle-node, 
at the same value of $L$. Figure~\ref{fig:ridge-prof-hori}(d) shows the 
(stable) drop state at $L=30>L_{c1}^0$. We emphasize that the notion of 
stability is limited to linear stability with respect to perturbations with 
spatial period $L$. It turns out that the large amplitude drop branch is 
typically unstable w.r.t.\ coarsening, i.e., to perturbations with periods that 
are integer multiples of $L$. However, we do not pursue questions related to 
coarsening in the present work.

Figure~\ref{fig:celmu0eps0,3} shows, in addition to the SD$_1$ states,
four additional branches of drop-like states. Of these the states labeled
SD$_2$ and SD$_3$ lie on a pair of unstable branches that are connected 
via a common saddle-node bifurcation at $L_\mathrm{sn2}\approx25.0$ but 
that are disconnected from the ridge states SR$_1$, SR$_2$ and SR$_3$ 
(Fig.~\ref{fig:celmu0eps0,3}). The states labeled SD$_2$ consist of drops 
sitting on the defect with a maximum on the defect while SD$_3$ consist 
of drops sitting on the defect with maxima on either side of the defect 
[Fig.~\ref{fig:ridge-prof-hori2,3}(a,b)], just as for the corresponding 
SR$_2$ and SR$_3$ states. In addition, we find two further branches of 
drop-like states, hereafter SD$_4$ and SD$_5$, both created as a result 
of symmetry-breaking bifurcations of the SR$_2$ and SR$_3$ states.
Following \cite{HiKn98} we refer to these states as varicose and zigzag
states. The former are shown in Figs.~\ref{fig:ridge-prof-hori2,3}(c,d)
at two locations along the SD$_4$ branch; the latter solution is shown 
in Fig.~\ref{fig:ridge-prof-hori2,3}(e). As shown in Fig.~\ref{fig:celmu0eps0,3} 
the varicose [zigzag] branch bifurcates subcritically from SR$_2$ at 
$L_{c2}^0=19.95$ [$L^0_{c4}=34.52$] and connects to SR$_3$ supercritically at 
$L_{c3}^0=16.05$ [$L^0_{c5}=23.19$]. Note that the SD$_4$ branch is twice unstable
while the SD$_5$ branch is four times unstable. As explained in \cite{HiKn98}
this is a consequence of the fact that the zigzag state is unstable with respect
to two different varicose modes. Technically, the unstable eigenvalue
of double multiplicity splits into two eigenvalues when moving from
the SR branch onto the SD$_5$ branch.
%

\subsection{Ridge on an inclined substrate: $\mu>0$}
\mylab{sec:incl-ridge}
%
Once the driving force $\mu$ becomes nonzero, the ridge or drop solutions 
will be displaced from their symmetric location midway between a pair of 
hydrophobic defects. This is a consequence of the term $-\mu\partial_xQ(h)$ 
in Eq.~(\ref{eq:lub}) that breaks the reflection symmetry in the 
$x$-direction. On a homogeneous substrate the solutions
acquire an asymmetric shape w.r.t.\ $x\rightarrow -x$ and slide downstream
like a solitary wave of constant shape~\cite{Thie01,ThKN04}. On a
heterogeneous substrate, however, the wettability defect may pin such a 
sliding solution depending on the relative size of the parameters $\mu$ and
$\epsilon$. In the present case of a hydrophobic defect with $\epsilon=0.3$
the ridge or drop will be blocked at its front. A finite force 
$\mu>\mu_\mathrm{depin}$ is needed to overcome the pinning effect of the 
defect in order that the ridge/drop may move. Beyond this depinning 
transition ridges/drops slide along the plane but do so with a nonuniform 
speed since both their shape and speed are modulated
periodically as they pass individual defects in the periodic defect array
\cite{ThKn06,ThKn06b,BHT09}. In 2d the depinning itself is either related 
to a saddle-node infinite period (sniper) bifurcation or to a Hopf
bifurcation. As already mentioned the sniper bifurcation generates stick-slip
motion just above $\mu_\mathrm{depin}$ since the ridge sticks for a long 
time to a defect before sliding to the next one \cite{ThKn06b}.

In this section we seek to elucidate the depinning process for 3d drops
and relate it to the corresponding process in 2d. We begin with stability
and depinning of the 2d states SR.

\subsubsection{Depinning in two dimensions}
\mylab{incl-ridge-inv}
%
%
\begin{figure}[tbh]
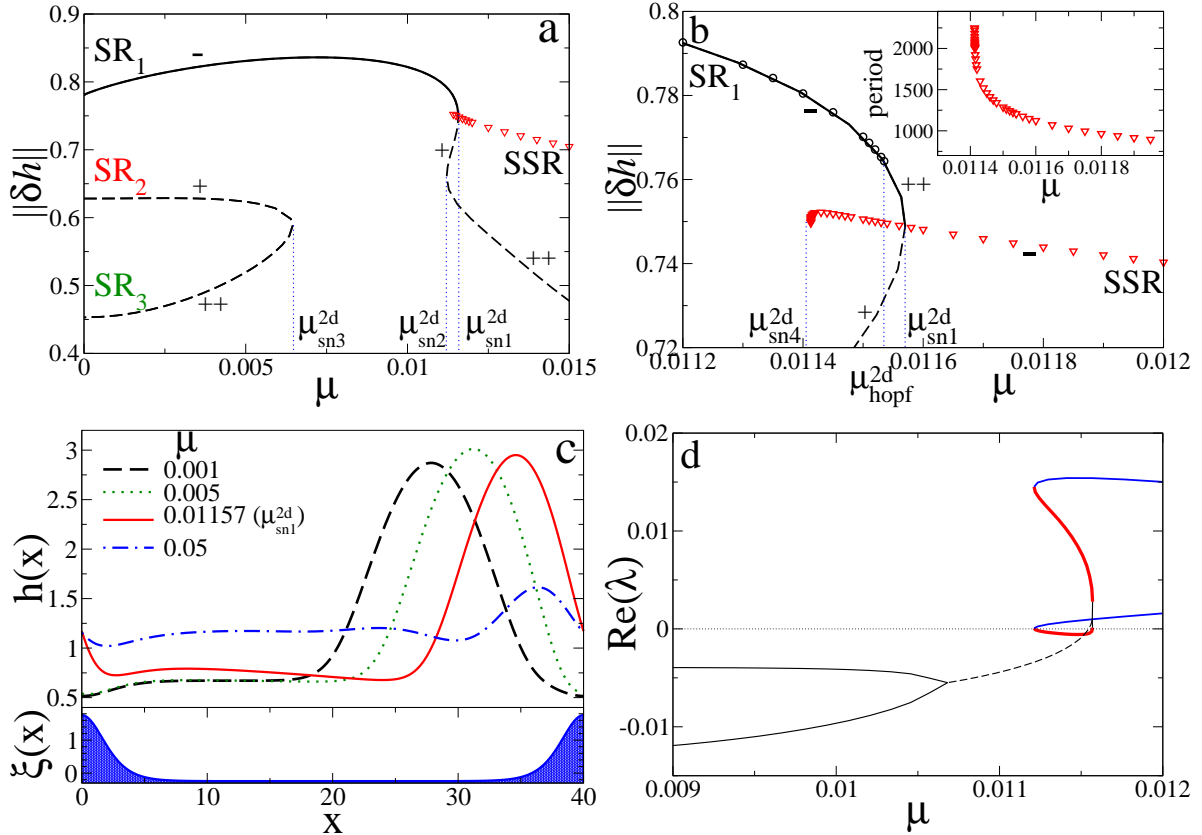

\centering
\includegraphics[width=0.475\hsize]{fig7a}
\includegraphics[width=0.475\hsize]{fig7b}\\[1ex]
\includegraphics[width=0.475\hsize]{fig7c}
\includegraphics[width=0.475\hsize]{fig7d}
\caption%
{(color online) (a) Bifurcation diagram showing the amplitude of the three steady 2d states
identified in Fig.~\ref{fig:ridge-diag-prof-hori} as a function of $\mu$.
Solid [dashed] lines indicate solutions that are stable [unstable] with respect
to perturbations with a real eigenvalue. At larger $\mu$ only SR$_1$ exits. 
(b) Zoom of (a) showing in addition the branch of stable stick-slip 
states (SSR) (open triangles); the inset shows the time taken to travel between 
successive defects, also as a function of $\mu$. The open circles denotes
solutions that are stable with respect to oscillations. The symbols $\pm$ 
indicate the stability of the branches w.r.t.\ 2d perturbations, with $-$ 
indicating stability and the $+$'s indicating the number of unstable eigenmodes.
(c) Solution profiles along the upper SR$_1$ branch showing the blocking of the ridge 
by the downstream defect (the driving force acts towards the right). (d) The real 
part of the leading eigenvalues of the SR$_1$ states as a function of $\mu$.
The solid [dashed] lines correspond to real [complex] eigenvalues. The three 
different line widths correspond to the three sub-branches separated by 
saddle-node bifurcations. Parameters: $\epsilon=0.3$, ${\bar h}=1.2$, $L_x=40$.
}
\mylab{fig:cpnob2d}
\end{figure}

Figure~\ref{fig:cpnob2d}(a) presents the bifurcation diagram for $\epsilon=0.3$ 
(hydrophobic defects) as a function of the driving force $\mu$ obtained using 
numerical continuation and direct numerical simulation in time. The figure shows 
2d states only, with the $L^2$ norm (\ref{eq:norm}) for steady ridges (SR) and
the time-averaged $L^2$ norm (\ref{eq:tnorm}) for the stick-slip states (SSR). 
As soon as $\mu\ne0$
the profiles of the SR states become markedly asymmetric as shown for SR$_1$ in 
Fig.~\ref{fig:cpnob2d}(c). The figure illustrates the dramatic shift in the 
position of the dominant maximum of the solution towards the downstream defect 
arising from the competition between the driving force and the blocking effect 
of the defect that prevents downstream motion. As the driving force $\mu$ 
increases the ridge profile steepens (i.e., the norm (\ref{eq:norm}) increases).
However, close to the saddle-node bifurcation at 
$\mu=\mu_\mathrm{sn1}^\mathrm{2d}\approx 0.01157$ the norm starts to decrease, 
and for the parameters used the saddle-node bifurcation is preceded by a Hopf
bifurcation at $\mu_\mathrm{hopf}^\mathrm{2d}\approx 0.01154$ 
[Fig.~\ref{fig:cpnob2d}(c)]. This bifurcation leads to a subcritical branch of 
unstable small amplitude time-periodic ``rocking'' states hereafter referred to as oscillating ridges (OR). We conjecture that
with decreasing $\mu$ these states undergo a global bifurcation to translating 
states at some $\mu=\mu_{g}^\mathrm{2d}$ 
($\mu_{g}^\mathrm{2d}<\mu_\mathrm{hopf}^\mathrm{2d}<\mu_\mathrm{sn1}^\mathrm{2d}$)
These states differ, however, from the usual stick-slip states generated via the 
sniper bifurcation since the ridge must spend considerable time in the rocking state 
before shifting rapidly to the next defect downstream.

\begin{figure}[tbh]
\centering
{\large (a)}\includegraphics[width=0.45\hsize]{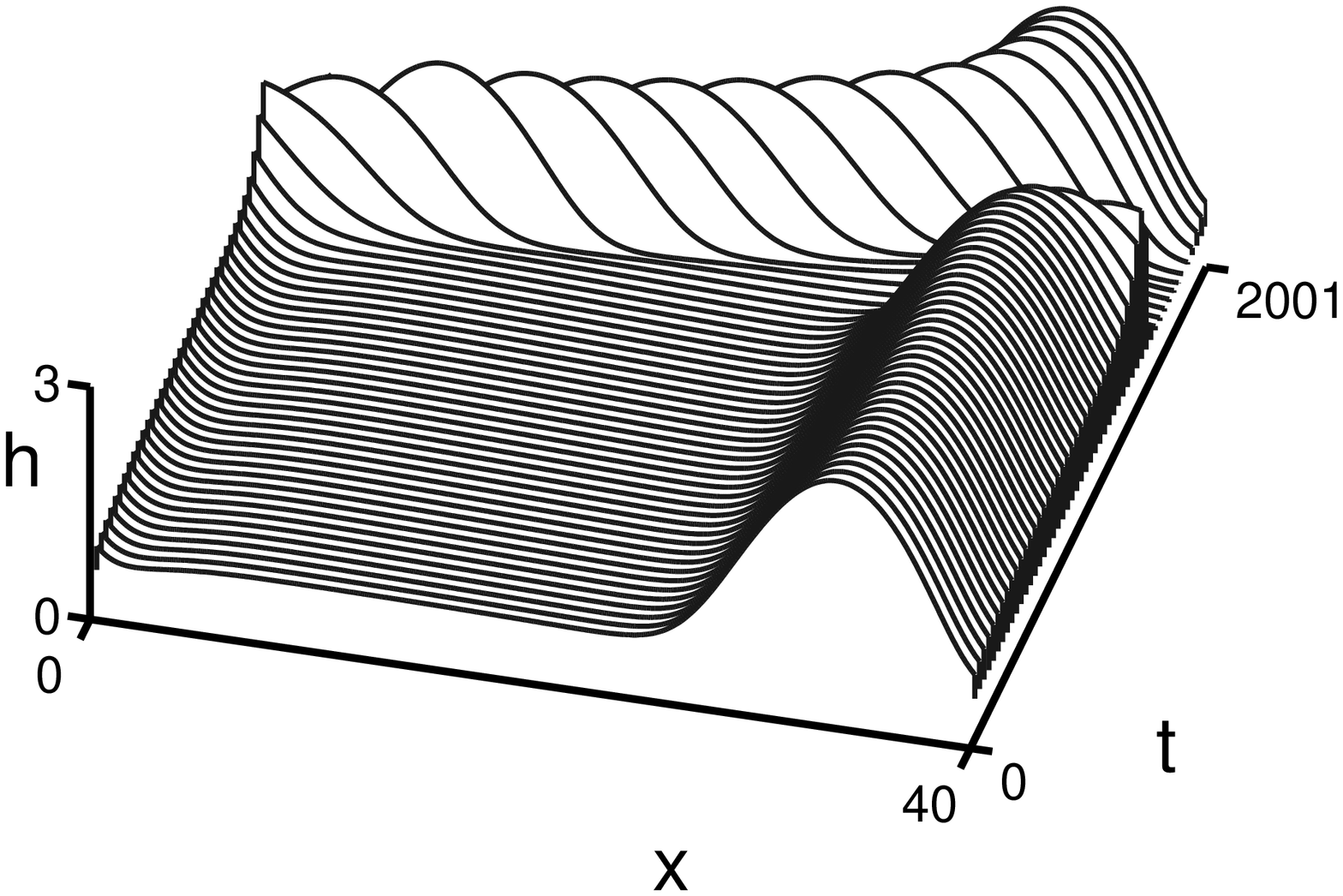}
\includegraphics[width=0.45\hsize]{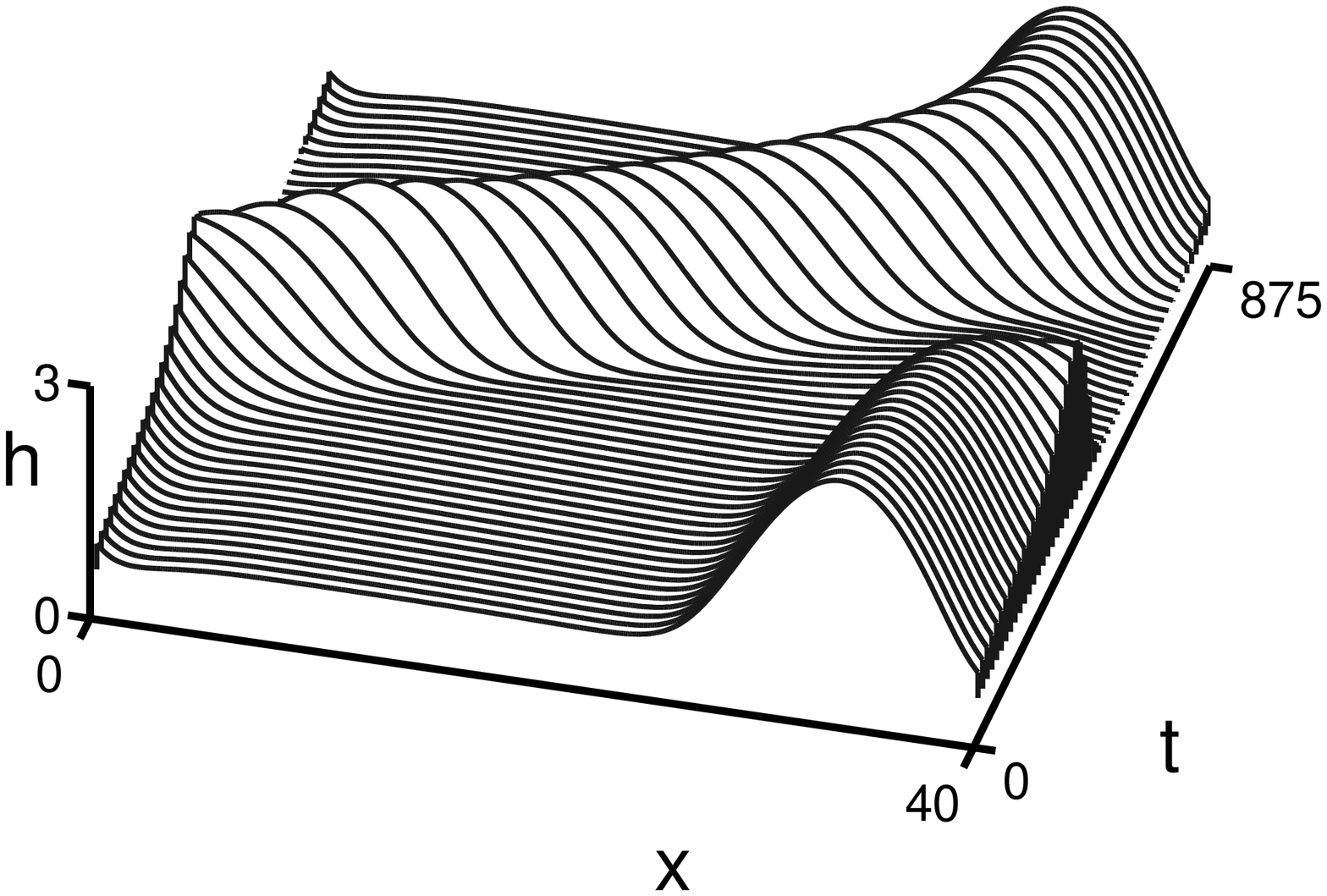}{\large (b)}\\
 {\large (c)}\includegraphics[width=0.45\hsize]{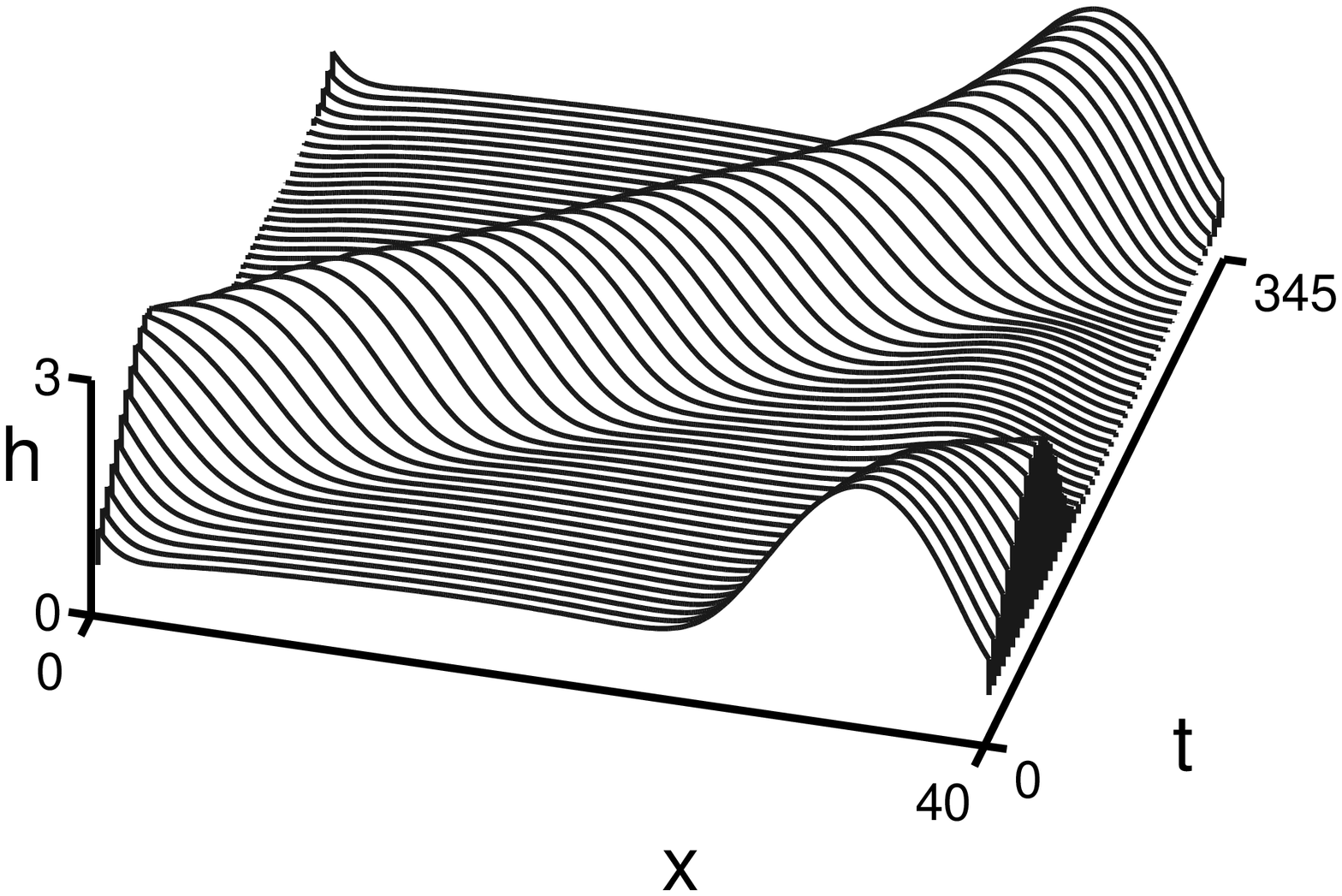}
\includegraphics[width=0.45\hsize]{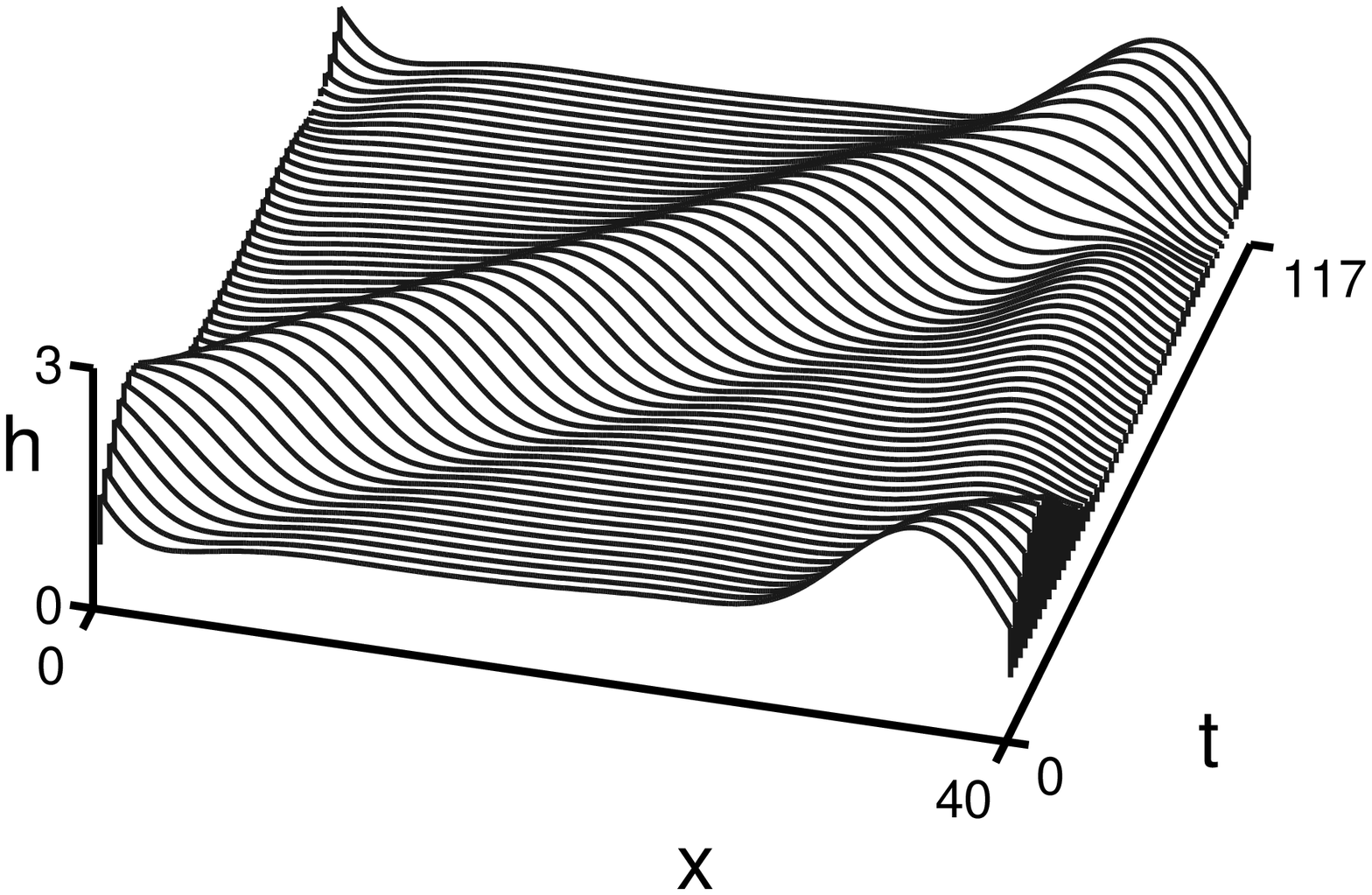}{\large (d)}
\caption{Space-time plots of the time evolution of
  stick-slipping ridge (SSR) solutions beyond depinning via a Hopf
  bifurcation over one period in space and time at (a)
  $\mu=0.011414$ close to the saddle-node of the SSR branch at
  $\mu_\mathrm{sn4}^\mathrm{2d}=0.01141303$ with temporal period 
  $T=2000.6$, (b) $\mu=0.012$ with $T=875.0$, (c) $\mu=0.02$
  with $T=345.0$, and (d) far from depinning, $\mu=0.06$ with $T=116.8$. }%
\mylab{fig:spti-2d}
\end{figure}

\begin{figure}[tbh]
\centering
\includegraphics[width=0.7\hsize]{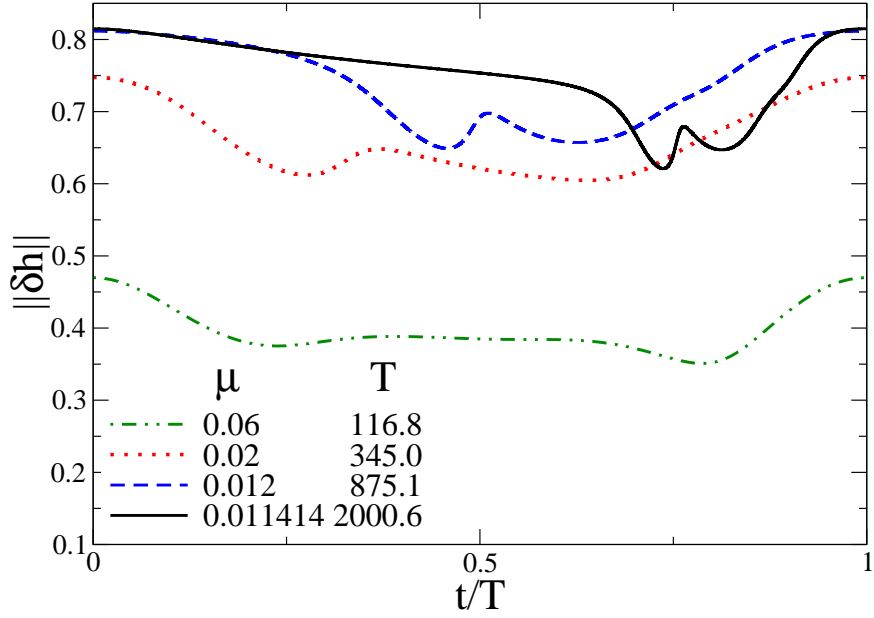}
\caption{(color online) The $L^2$ norm as a function of time for selected
solutions on the branch of stick-slip ridges (SSR). Time is scaled by
the period $T$ listed in the legend together with the corresponding
values of $\mu$.}%
\mylab{fig:SSR-norm}
\end{figure}

Figure~\ref{fig:cpnob2d}(b) zooms into the region of the bifurcation
diagram where stable stick-slip ridges (SSR) are present, while
Figs.~\ref{fig:spti-2d} and \ref{fig:SSR-norm} show several of these
states as a function of time over one period $T$ in terms of
space-time plots (Fig.~\ref{fig:spti-2d}) and $L^2$ norm
(Fig.~\ref{fig:SSR-norm}). Here the 'period' refers to the time
required for the ridge to slide from one defect to the
next. Figure~\ref{fig:cpnob2d}(b, inset) suggests that the period $T$
diverges logarithmically as $\mu\rightarrow\mu_{g}$ as
expected from a global bifurcation. Examination of the SSR profile
near $\mu=\mu_{g}$ when it is almost pinned suggests that
the SSR approaches close to an SR$_1$ on the intermediate segment
of the SR$_1$ branch (not shown). The leading eigenvalues along the 
SR$_1$ branch are shown in Fig.~\ref{fig:cpnob2d}(d) with the heavy red 
lines indicating the eigenvalues along the intermediate segment.
For $\mu\approx \mu_{g}$ the states on this segment have a
single unstable eigenvalue $\lambda_u\approx0.0101$ and a leading
stable eigenvalue $\lambda_s\approx -0.00054$. Thus $\lambda_u+\lambda_s>0$ 
and under these conditions standard theory shows that any global 
bifurcation involving states on the intermediate segment of the 
SR$_1$ branch must involve {\it unstable} periodic states (see the 
3d case below for further discussion). This conclusion is consistent 
with the behavior of the SSR norm [Fig.~\ref{fig:cpnob2d}(b)] which
suggests that the SSR states do indeed undergo a saddle-node
bifurcation prior to any global bifurcation and so are unstable near
$\mu=\mu_{g}$. 

The leading eigenvalues on the intermediate segment of SR$_1$ are created
by the collision on the positive real axis of the unstable complex 
eigenvalues responsible for the Hopf bifurcation on the upper segment 
of the SR$_1$ branch. After this collision the eigenvalues remain real
with one continuing to increase as one follows the branch while the 
other crosses back into the negative half plane at the saddle-node 
$\mu_\mathrm{sn1}^\mathrm{2d}$. At this point the steady branch turns back 
towards smaller $\mu$ [Fig.~\ref{fig:cpnob2d}(b)]. The bifurcation at 
$\mu_\mathrm{sn1}^\mathrm{2d}$ is a standard saddle-node bifurcation and not a 
sniper. This is because for the parameter values used the depinning transition 
involves the unstable OR states instead of the SR$_1$ states, i.e., with the 
appearance of the Hopf bifurcation the depinning transition moves from the 
steady states SR$_1$ to the time-dependent OR states. At a further 
saddle-node at $\mu_\mathrm{sn2}^\mathrm{2d}\approx 0.0112$ the branch turns again 
towards increasing $\mu$, now with two unstable real eigenvalues 
[Fig.~\ref{fig:cpnob2d}(c)]. With further increase in $\mu$ the $L^2$ norm 
of these unstable states continues to decrease as the solution begins to 
resemble more and more a 2d steady flowing film whose profile is modulated 
by the hydrophobic defects in the substrate below.

Since each of the three SR states at $\mu=0$ can itself be continued in $\mu$ 
we expect to find additional SR states when $\mu>0$. Of these SR$_2$ and SR$_3$ 
annihilate in a saddle-node bifurcation at $\mu_\mathrm{sn3}^\mathrm{2d}\approx 0.0065$
[Fig.~\ref{fig:cpnob2d}(a)] leaving only the state SR$_1$ at larger $\mu$. 
These states play no role in the depinning transition and the bifurcation 
at $\mu_\mathrm{sn3}^\mathrm{2d}$ is a regular saddle-node bifurcation.

For larger ridge volumes (e.g., ${\bar h}=1.3$) the bifurcation diagram 
simplifies since the Hopf bifurcation is now absent. In this case the branch 
of stick-slip ridges (SSR) emerges directly from the saddle-node bifurcation 
in a sniper bifurcation, as confirmed by the fact that the inverse time-period 
now varies as $(\mu-\mu_\mathrm{sn1}^\mathrm{2d})^{1/2}$ near the bifurcation. 
For a more detailed discussion of the transition between these two scenarios 
see Ref.~\cite{ThKn06}.

\subsubsection{Transverse instability of a ridge} \mylab{sec:incl2s}
A real 3d system will, however, show the behavior described in the
previous section only if the spanwise domain size $L$ is
small. For large or indeed infinite $L$ the ridge solutions (2d drops)
are unstable, as already noted, with respect to spanwise perturbations.
On a horizontal substrate ($\mu=0$) the only relevant instability of this type 
is the Plateau-Rayleigh instability discussed in section~\ref{sec:hori}. 
In this section we discuss analogous instabilities when $\mu>0$.

To do so we follow the SR$_1$ branch in Fig.~\ref{fig:cpnob2d}(a) as a function
of $\mu$ together with the eigenvalues determining its linear stability
properties with respect to spanwise perturbations of spatial period $L$.
This procedure allows us to identify, as a function of $\mu$, the critical 
spanwise domain size $L_c$ for the onset of a linear instability.  
The underlying linear stability analysis is performed using the Ansatz 
$h(x,y)=h_0(x)+\alpha h_1(x)\exp(ik_yy)$ with $k_y=2\pi/L$.  

The resulting stability diagram in the $(L,\mu)$ plane is
shown in Fig.~\ref{fig:cpelob}. Figure~\ref{fig:eig3d} shows four
sample perturbation profiles $h_1(x)$ along the stability boundary
together with the associated ridge profiles $h_0(x)$. These are discussed below.

\begin{figure}[tbh]
\centering
\includegraphics[width=0.7\hsize]{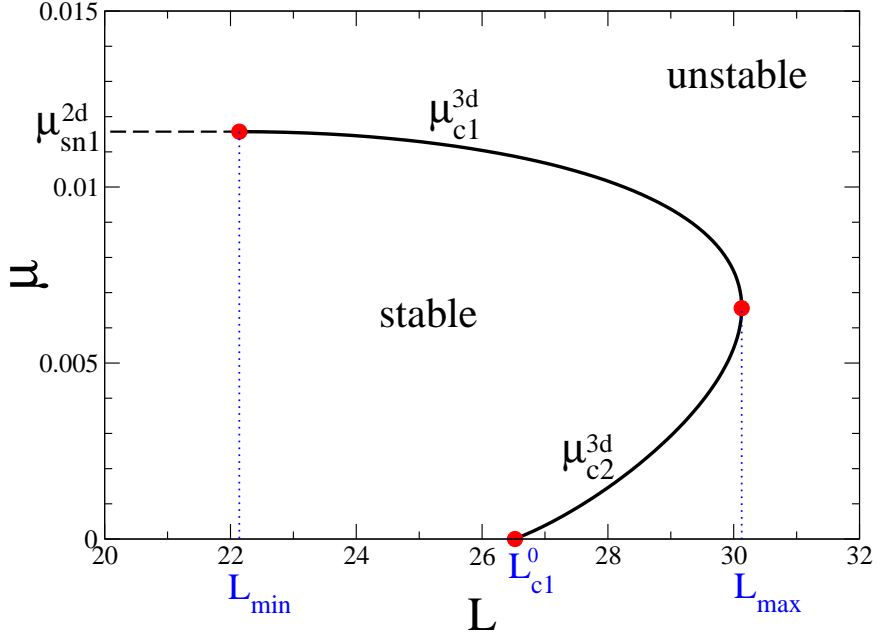}
\caption%
{(color online) Stability diagram for steady ridge states on a heterogeneous incline showing 
the region of linear stability in the $(L,\mu)$ plane. Solid (dashed) lines
indicate the presence of 3d (2d) instability. The remaining parameters 
are as in \myfig{cpnob2d}. }%
\mylab{fig:cpelob}
\end{figure}

\begin{figure}[tbh]
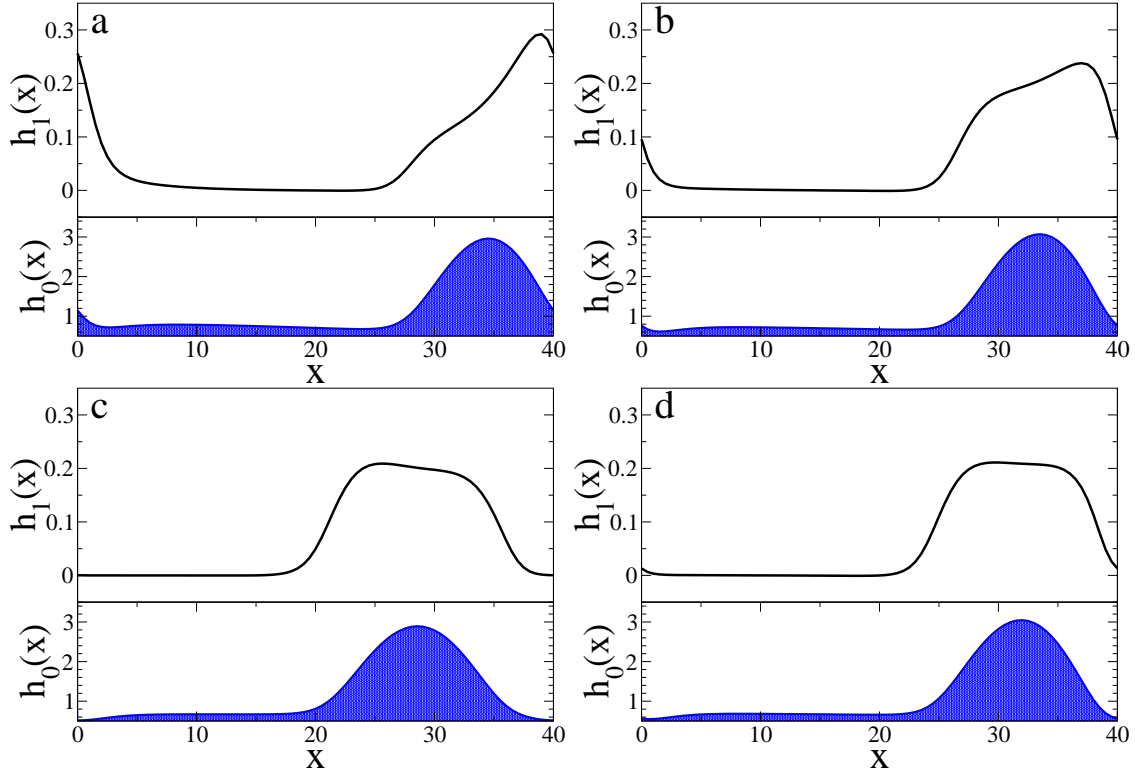

\centering
\includegraphics[width=0.45\hsize]{fig11a}
\includegraphics[width=0.45\hsize]{fig11b}\\
\includegraphics[width=0.45\hsize]{fig11c}
\includegraphics[width=0.45\hsize]{fig11d}
\caption%
{(color online) The marginal eigenfunctions $h_1(x)$ along the stability boundary in
Fig.~\ref{fig:cpelob} (upper panels) and associated steady ridge
profiles $h_0(x)$ (lower panels) at (a) $\mu=0.01157$, 
$L=22.5$; (b) $\mu=0.01019$, $L=28.0$; (c) $\mu=0.00147$, $L=28.0$; 
(d) $\mu=0.00656$, $L_\mathrm{max}=30.1$.
}
\mylab{fig:eig3d}
\end{figure}

On the horizontal substrate ($\mu=0$) the SR$_1$ states are
stable below the critical length $L_{c1}^0$. When $\mu>0$ these
states become asymmetric with respect to the reflection
$x\rightarrow-x$ [Fig.~\ref{fig:cpnob2d}(b)] but remain stable for
$\mu<\mu_{c1}^{3d}(L)$. For $L_{c1}^0<L<L_\mathrm{max}$ the SR$_1$ states are 
stable or unstable depending on the value of $\mu$. This is a consequence 
of the nonmonotonic dependence of $L_{c1}$ on $\mu$ shown in
Fig.~\ref{fig:cpelob} for $\epsilon=0.3$: the value of $L_{c1}$
first increases from $L_{c1}^0$ to a maximum value $L_\mathrm{max}$
at $\mu_{\mathrm{max}}$ before decreasing towards $L_\mathrm{min}$ at 
$\mu_\mathrm{sn1}^\mathrm{2d}$. For the present parameter values 
$L_\mathrm{min}<L_{c1}^0$ and linear considerations alone allow one 
to distinguish four qualitatively different responses to the driving 
force $\mu$, depending on the lateral system size $L$:
\begin{itemize}
\item[(i)] For $L\leq L_\mathrm{min}= 22.4$ the pinned ridge
  is linearly stable with respect to spanwise perturbations in its 
  entire range of existence, i.e., for all
  $\mu<\mu_\mathrm{sn1}^\mathrm{2d}$. The depinning
  behavior corresponds to the 2d case.
\item[(ii)] For $L_{\mathrm{min}}<L<L_{c1}^0$ the ridge is linearly
  stable when $\mu=0$ but loses stability at a driving force 
  $\mu_{c1}^\mathrm{3d}<\mu_\mathrm{sn1}^\mathrm{2d}$. The instability
  has largest amplitude on the downstream side of the ridge
  [Fig.~\ref{fig:eig3d}(a)] indicating the onset of spanwise modulations 
  that invade the wettability defect.  As $L$ approaches
  $L_\mathrm{min}$, $\mu_{c1}^\mathrm{3d}$ approaches 
  $\mu_\mathrm{sn1}^\mathrm{2d}$. 
\item[(iii)] For $L_{c1}^0<L<L_\mathrm{max}$ the SR$_1$ state is
  linearly unstable even when $\mu=0$. Increasing $\mu$ stabilizes the 
  ridge at a critical value $\mu_{c2}^\mathrm{3d}$. The marginal
  eigenfunction [Fig.~\ref{fig:eig3d}(c)] peaks slightly on the upstream
  side of the ridge indicating stabilization with respect to the
  Plateau-Rayleigh instability. The ridge is then linearly 
  stable up to the critical value $\mu_{c1}^\mathrm{3d}$, where it
  loses stability as in case (ii), cf. Fig.~\ref{fig:eig3d}(a,b).
\item[(iv)] For $L>L_\mathrm{max}= 30.7$ the ridge is linearly
  unstable for all $\mu$. However, linear analysis is not able to tell whether
the ridge will evolve into steady pinned drops or sliding drops. We expect
that a critical $\mu$ exists below [above] which the former [latter] occurs.
\end{itemize}

The different types (i)--(iv) of linear behavior inevitably result in
different nonlinear behavior. For example, in cases (ii) and (iii)
depinning occurs in an intrinsically 3d manner, as individual
``fingers'' extend across the defect. Thus the SR$_1$ state does not
slide over the defect as a whole.  In the following section we analyse
the relation between drop and ridge solutions in the parameter regions
(i) to (iv). We find that the intricate nonlinear behavior that
results requires the introduction of several qualitatively different
subregions in parameter space. These are discussed in detail in the
next Section before an overview of all stable solutions is given in
Section~\ref{sec:all-stab}.
%
\subsection{Depinning of drops and ridges on an inclined substrate} 
\mylab{sec:dropridge}
%
\begin{figure}[tbh]
\centering
\includegraphics[width=0.7\hsize]{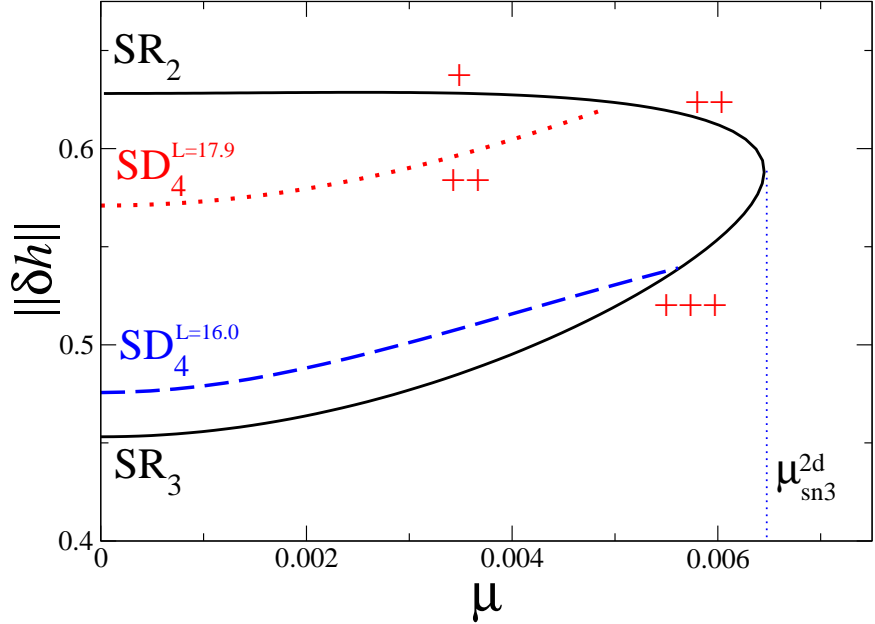}
\caption%
{(color online) Bifurcation diagram showing the $L^2$ norm
    $||\delta h||$ of steady solutions SR$_2$ and SR$_3$ as a function 
    of the driving force $\mu$ together with the bifurcating
    transversally modulated ridges SD$_4$. The latter branch is shown
    for the two different values $L=17.9$ (red dotted) and $L=16.0$
    (blue dashed). The number of red $+$'s indicates the number of 
    unstable eigenmodes when $L=17.9$.  Remaining parameters are as in
    Fig.~\ref{fig:ridge-diag-prof-hori}.}
  \mylab{fig:cpnelmin}
\end{figure}

The analysis of the linear stability of the steady ridge solutions for
different spanwise system sizes performed in the previous section
indicates that the relation between drop and ridge solutions changes
with lateral system size, i.e., with drop volume. The linear analysis
allowed us to distinguish four parameter regimes. The curve of
neutral stability w.r.t.\ harmonic spanwise perturbations in
Fig.~\ref{fig:cpelob} indicates the loci of bifurcation points where
modulated ridge solutions emerge from the spanwise-invariant
ridges. In the following, we continue these modulated ridge solutions 
for several fixed values of $L$ while changing $\mu$. In addition,
we continue the steady drop-like states present at $\mu=0$ 
(section~\ref{sec:hori}, Fig.~\ref{fig:celmu0eps0,3}) towards larger 
$\mu$. All solutions obtained in this way are presented in a sequence of
bifurcation diagrams together with the spanwise-invariant ridge
solutions. These show branches of time-periodic solutions, i.e., drops
and ridges that either oscillate or slide from defect to defect, in 
addition to the steady solutions.

\subsubsection{Scenario (i): $L\leq L_\mathrm{min}$}
For $L\leq L_\mathrm{min}$ the pinned ridge is linearly stable w.r.t.\
spanwise perturbations and depins via a Hopf bifurcation at
$\mu_\mathrm{hopf}^\mathrm{2d}<\mu_\mathrm{sn}^\mathrm{2d}$.  The
marginally stable eigenfunction at the Hopf bifurcation remains
translation-invariant in the spanwise direction. Consequently the
depinning scenario in this case is identical to that already described
for the 2d case in section~\ref{incl-ridge-inv} and summarized in
Fig.~\ref{fig:cpnob2d}. After depinning the ridge undergoes stick-slip
motion as described for 2d drops in
Refs.~\cite{ThKn06,ThKn06b}. Depending on the value of $L\leq
L_\mathrm{min}$ this 2d time-dependent state may in turn become
unstable to three-dimensional perturbations at some
$\mu>\mu_\mathrm{sn4}^\mathrm{2d}$, resulting in a sliding 3d state.
  
The secondary branch SD$_4$ present on the horizontal substrate for
$L_{c3}^0<L<L_{c2}^0<L_\mathrm{min}$ (Fig.~\ref{fig:celmu0eps0,3})
plays no role in any of the depinning scenarios discussed below.
Continuation in $\mu$ shows that this solution terminates either on
SR$_2$ or SR$_3$, depending on the value of $L$ (Fig.\ref{fig:cpnelmin}).
The SD$_5$ solution likewise plays no significant role despite its 
presence in most of the scenarios discussed below 
(cf.~Figs.~\ref{fig:cpnel26} and \ref{fig:cpnel32} below).

\subsubsection{Scenario (ii-a): $L_\mathrm{min}\leq L\leq L_\mathrm{sn1}^0$}
For $L_\mathrm{min}\leq L\leq L_{c1}^0$ the pinned ridge SR$_1$ is
linearly stable w.r.t.\ spanwise perturbations at small $\mu$ but
becomes unstable w.r.t.\ 3d perturbations at some $\mu_{c1}^\mathrm{3d}(L)<\mu_\mathrm{sn1}^\mathrm{2d}$.  
Thus the Hopf bifurcation that produces depinning may be preceded by a steady state
spanwise instability. In this case the depinning that takes place at larger $\mu$
corresponds to an instability of a 3d steady state.

We have already seen that when $\mu=0$ and $L<L_\mathrm{sn1}^0$ we have 
the three SR solutions shown in Fig.~\ref{fig:cpnob2d}. For 
$L_\mathrm{sn1}^0<L<L_{c1}^0$ we have in addition two 3d solutions,
one ridge-like and the other drop-like. Each of these solutions
extends smoothly to $\mu>0$ with both $L_\mathrm{sn1}$ and $L_{c1}$
changing with $\mu$, with $L_\mathrm{sn1}(\mu)<L_\mathrm{sn1}^0$.
Thus for $L_\mathrm{min}\leq L\leq L_\mathrm{sn1}^0$ the
bifurcation diagram takes the form shown in Fig.~\ref{fig:cpnel23}.

\begin{figure}[h]
\centering
\includegraphics[width=0.45\hsize]{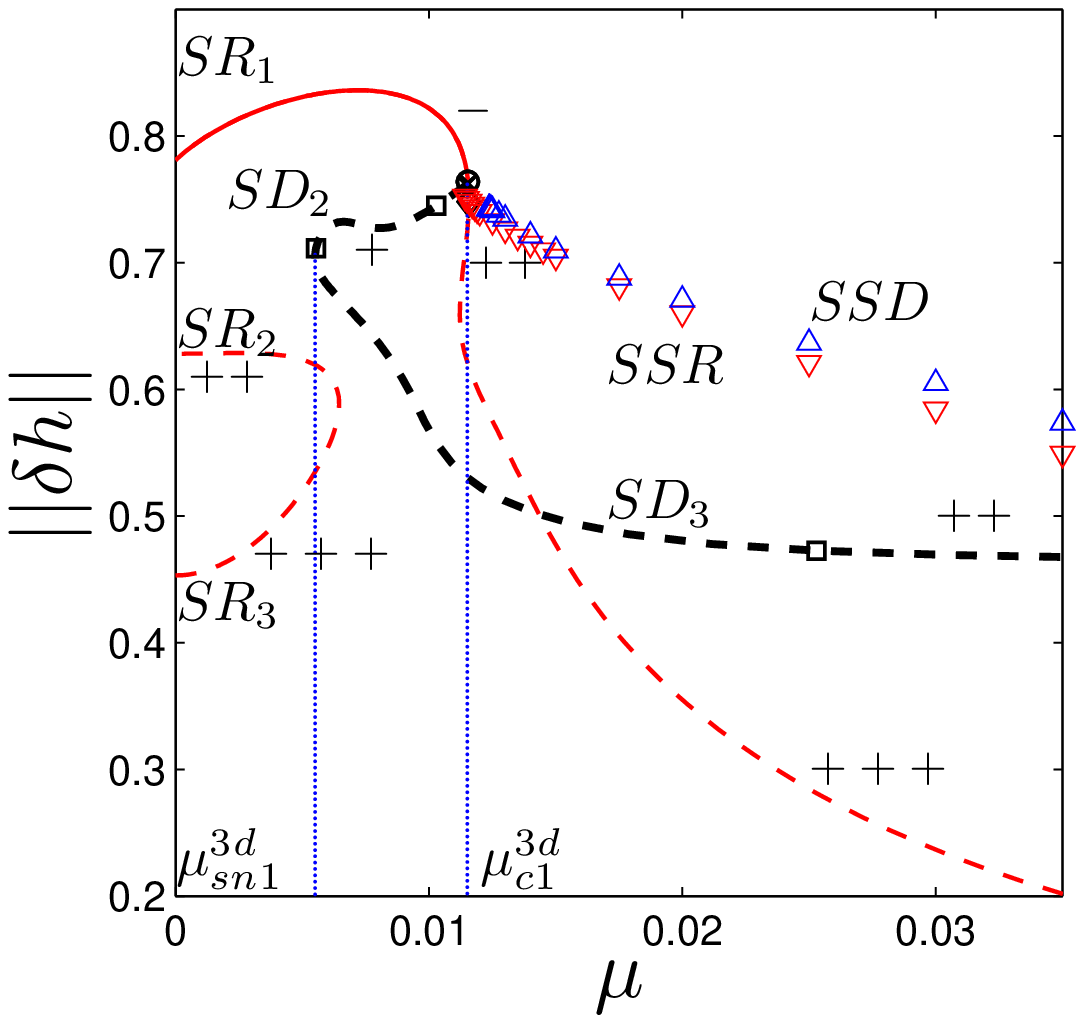}
\includegraphics[width=0.472\hsize]{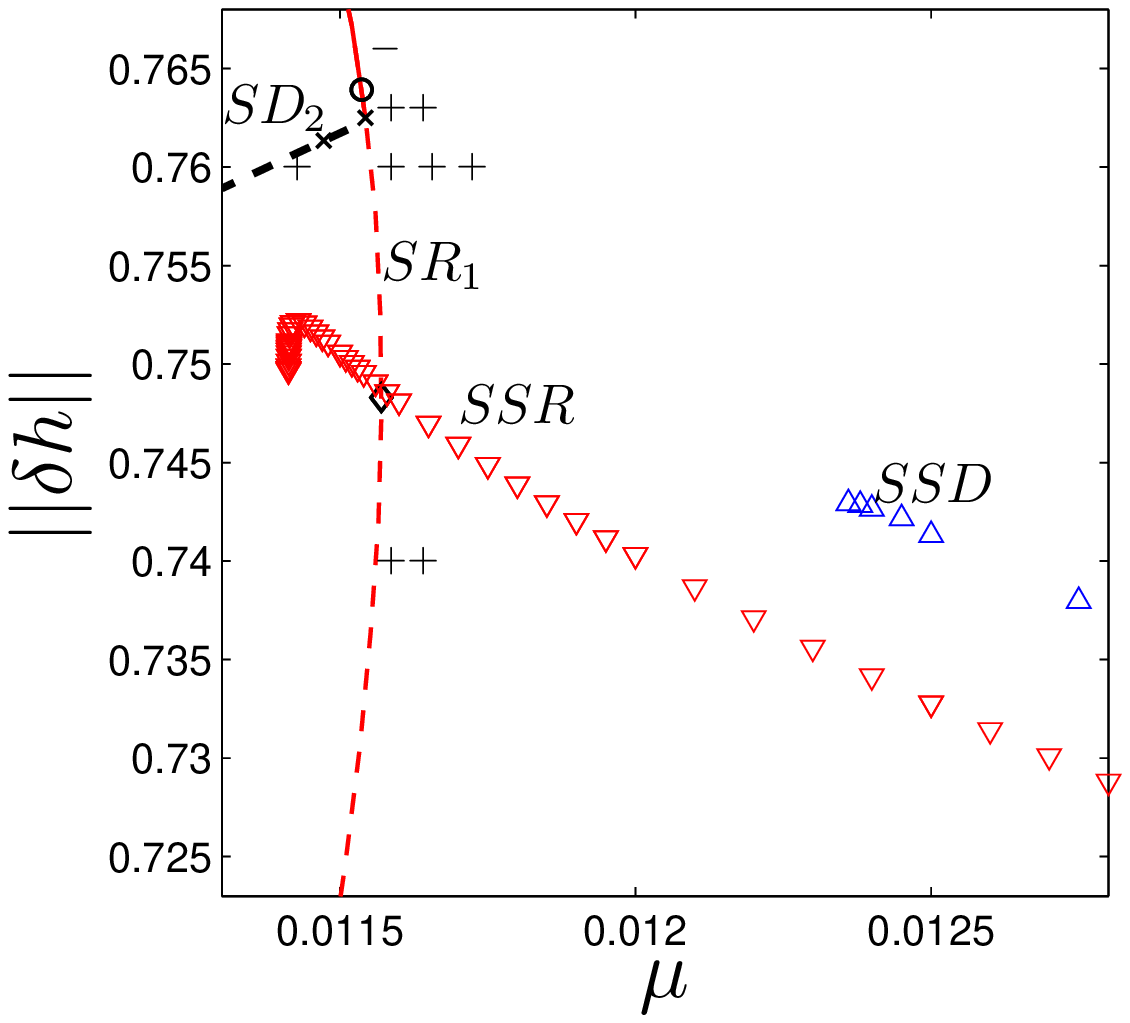}\\
{\large(a)\hfill(b)}
\caption%
{(color online) Scenario (ii-a): (a) Bifurcation diagram for ridge and
  drop solutions on a substrate with hydrophobic line defects
  (wettability contrast $\epsilon=0.3$) and spanwise system size
  $L_\mathrm{min}<L=23<L_\mathrm{sn1}^0$ showing the $L^2$ norm
  $||\delta h||$ of steady solutions as a function of the driving
  force $\mu$. A zoom is given in panel (b). Both steady spanwise
  invariant ridges (SR, thin red lines), and secondary drop solutions
  (SD$_2$ and SD$_3$, black lines) are included. Solid [dashed] lines
  indicate linearly stable [unstable] solutions. Downward [upward]
  pointing triangles indicate SSR [SSD] solutions. The former are
  taken from Fig.~\ref{fig:cpnob2d}(b) and are stable with respect to
  3d perturbations for $\mu<0.015$ only. Square symbols indicate
  solutions whose profiles are shown in Fig.~\ref{fig:snapel23}.  The
  SD$_3$ branch acquires stability at
  $\mu_\mathrm{hopf}^\mathrm{3d}\approx0.092$ (off scale).  The
  remaining parameters are as in \myfig{celnob}.  }
\mylab{fig:cpnel23}
\end{figure} 

\begin{figure}[h]
\centering
\includegraphics[width=0.8\hsize]{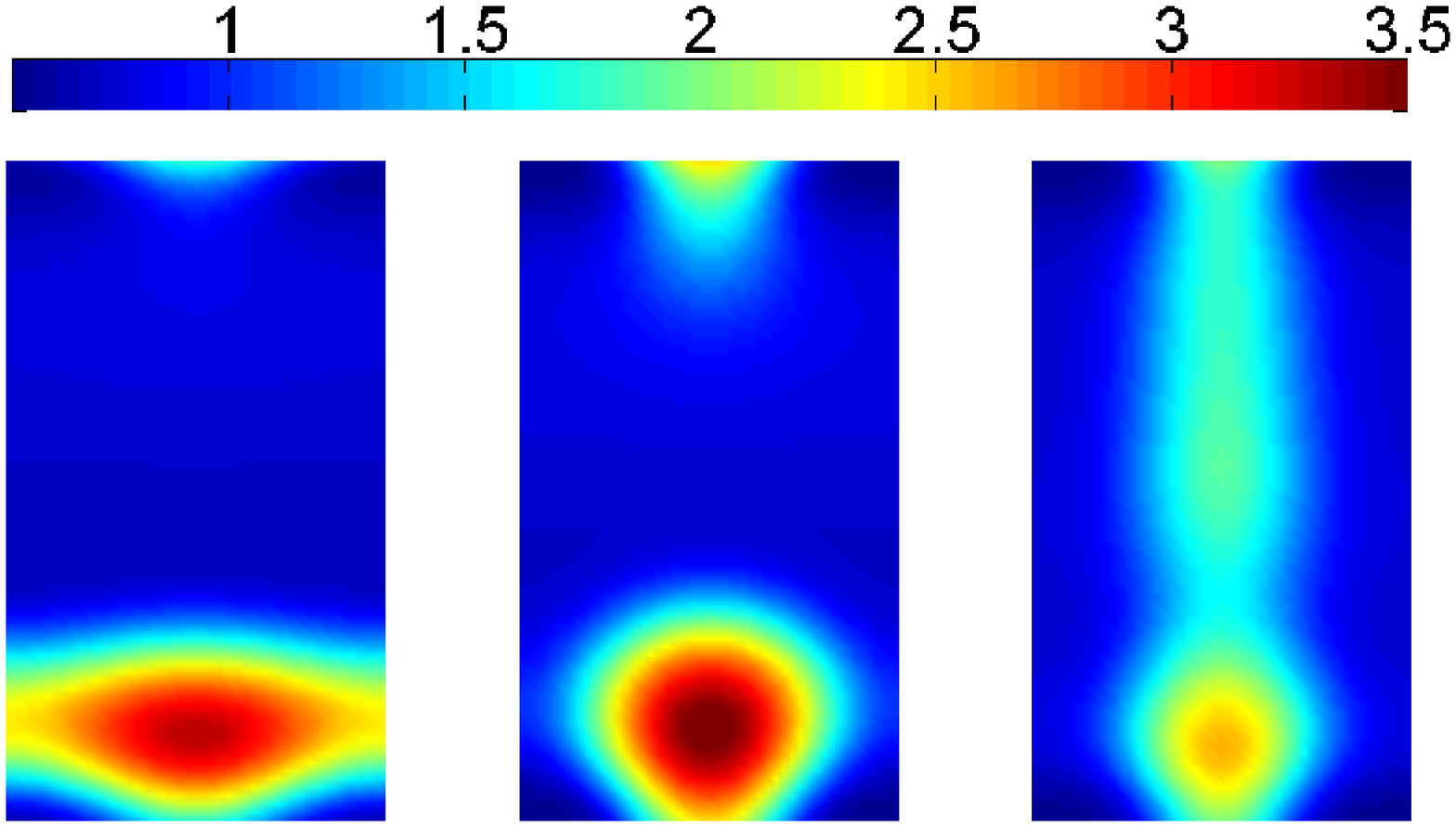}\\
{\large\hfill (a)\hfill(b)\hfill(c) \hfill\hspace*{.1cm}}
\caption%
{(color online) Scenario (ii-a): Unstable steady states at locations
  indicated by open squares in~\myfig{cpnel23} in terms of contours of 
  constant $h(x,y)$. (a) Drop solution SD$_2$ at $\mu=0.0105$
  resembling a modulated ridge. (b) SD$_2$ solution very close to
  the saddle-node at $\mu_{sn1}^\mathrm{3d}\approx 0.005532$. 
  (c) Static rivulet SD$_3$ at
  $\mu=0.0253$. Parameters are as in Fig.~\ref{fig:cpnel23}. }
\mylab{fig:snapel23}
\end{figure}

Here the branch of largest norm (thin line) corresponds to the steady
spanwise-invariant ridges SR$_1$. This solution loses stability
with respect to 3d perturbations at $\mu_{c1}^\mathrm{3d}\approx
0.01155$. This value is slightly larger than
$\mu_\mathrm{hopf}^\mathrm{2d}\approx 0.01154$ implying that the SR$_1$
state depins via 2d depinning. The 3d steady states produced at
$\mu_{c1}^\mathrm{3d}$ bifurcate towards smaller $\mu$ before turning 
around towards larger $\mu$ at a saddle-node bifurcation at
$\mu_\mathrm{sn1}^\mathrm{3d}\approx 0.005532$ and are unstable
throughout. In the following we use SD$_2$ to refer to the solutions
above the saddle-node and SD$_3$ to refer to those below. As shown 
in Fig.~\ref{fig:snapel23} the appearance of the 3d solutions changes 
substantially along the branch. Near $\mu_{c1}^\mathrm{3d}$
[Fig.~\ref{fig:snapel23}(a)] the solution is ridge-like. At the
saddle-node at $\mu_\mathrm{sn1}^\mathrm{3d}$ it becomes drop-like
[Fig.~\ref{fig:snapel23}(b)] while at large $\mu$ the solution becomes
a steady (unstable) streamwise-modulated rivulet
[Fig.~\ref{fig:snapel23}(c)]. 

As already mentioned the 2d SR$_1$ state loses stability at
$\mu=\mu_\mathrm{hopf}^\mathrm{2d}$ with respect to 2d oscillations
(see section \ref{sec:incl2s}, Fig.~\ref{fig:cpnob2d}) and these must
undergo a global bifurcation resulting in depinning (section
\ref{sec:int}). The resulting SSR state remains stable for a range of
values of $\mu$ but loses stability at $\mu\approx0.02$ to spanwise
perturbations resulting in a hysteretic transition to stable
stick-slip drops (SSD). The latter appear to undergo a saddle-node
bifurcation near $\mu=0.0123$ and do not reach any of the steady
states at lower $\mu$.

The unstable steady rivulet states on the SD$_3$ branch
[Fig.~\ref{fig:snapel23}(c)] stabilize through a Hopf bifurcation at
a much larger $\mu$, $\mu_\mathrm{hopf}^\mathrm{3d}\approx 0.092$.  
If a time simulation is done for $\mu<\mu_\mathrm{hopf}^\mathrm{3d}$ (but above
$\mu\approx0.01$) with the unstable steady rivulet solution as the
initial condition the rivulet decays into sliding drops
corresponding to a state on the SSD branch. Thus the SSD branch
terminates in a Hopf bifurcation on the SD$_3$ branch
(cf.~Fig.~\ref{fig:cpnel32} below).

\subsubsection{Scenario (ii-b): $L_\mathrm{sn1}^0 \leq L \leq L_c^0$}\label{sec:ii_b}

\begin{figure}[h]
\centering
\includegraphics[width=0.7\hsize]{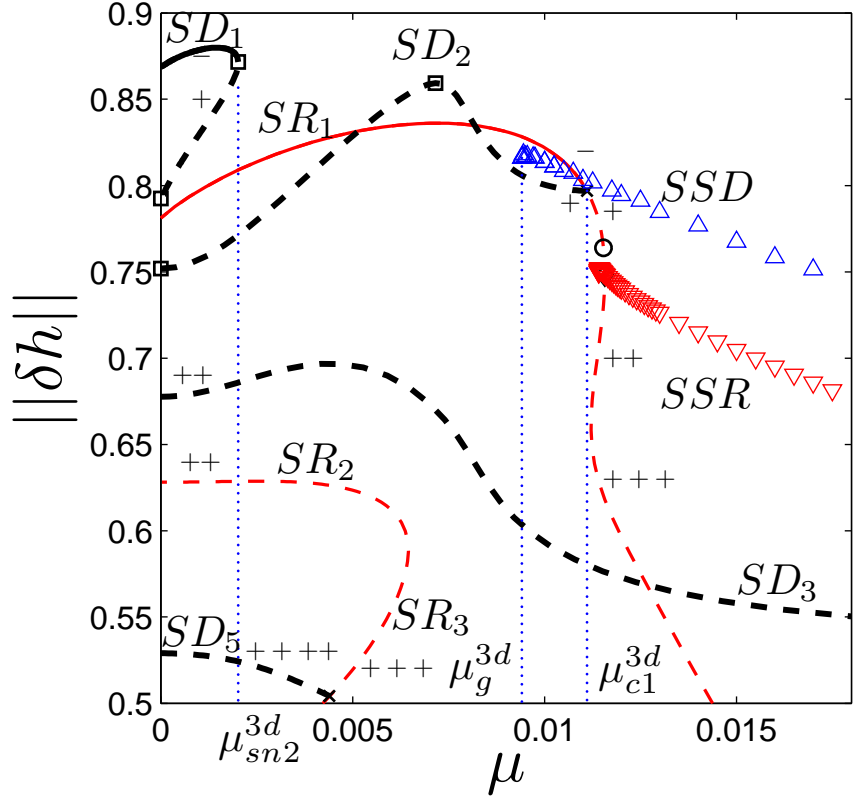}
\caption%
{(color online) Scenario (ii-b): Bifurcation diagram for ridge and
  drop solutions on a substrate with hydrophobic line defects
  (wettability contrast $\epsilon=0.3$) and spanwise system size
  $L=26$ showing the $L^2$ norm $||\delta h||$ of steady solutions 
  as a function of the 
  driving force $\mu$. Both steady spanwise invariant ridges 
  (SR$_1$, thin red lines), and secondary drop solutions (SD$_1$,
  SD$_2$, SD$_3$ and  SD$_5$, 
  black lines) are included. Solid [dashed] lines indicate
  linearly stable [unstable] solutions. Downward [upward] pointing
  triangles indicate SSR [SSD] solutions. Solution profiles at locations
  indicated by open squares are shown in Fig.~\ref{fig:snapel26}.
  The remaining parameters are as in \myfig{celnob}.  }
\mylab{fig:cpnel26}
\end{figure}

\begin{figure}[h]
\centering
\includegraphics[width=0.7\hsize]{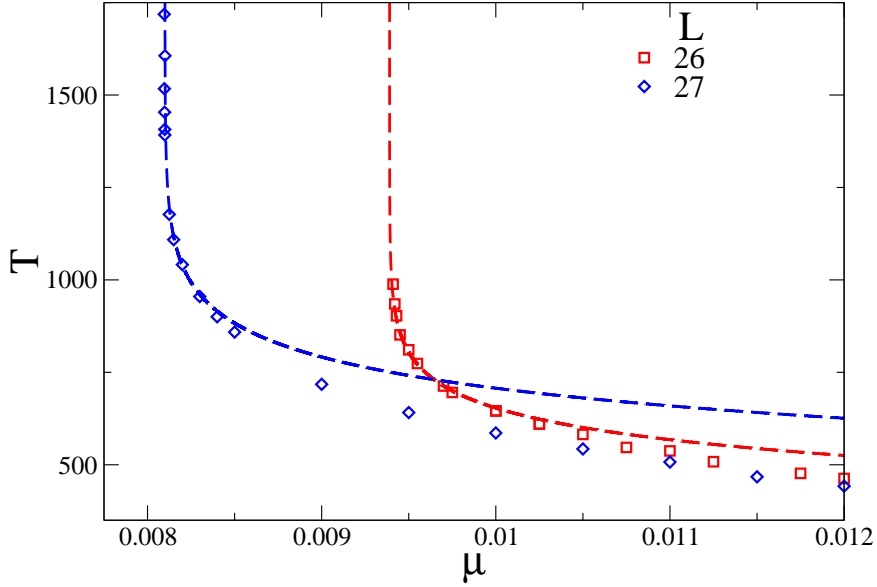}
\caption%
{(color online) Temporal period $T$ of stick-slipping drop (SSD) solutions as a function 
of the driving force $\mu$ for $L=26$ and $L=27$ (see legend). In each case the 
first six points were used for a fit of the form $T=-a_0\ln(\mu-a_1)$ (dashed 
lines). The fit parameters are $a_0=88.3$, $a_1=0.00939$ for $L=26$, and $a_0=112.85$,
  $a_1=0.00810$ for $L=27$.  }%
\mylab{fig:period-fits-SSD}
\end{figure}

\begin{figure}[h]
\centering
{\large(a)}\includegraphics[width=0.6\hsize]{fig17a}\\
{\large (b)}\includegraphics[width=0.6\hsize]{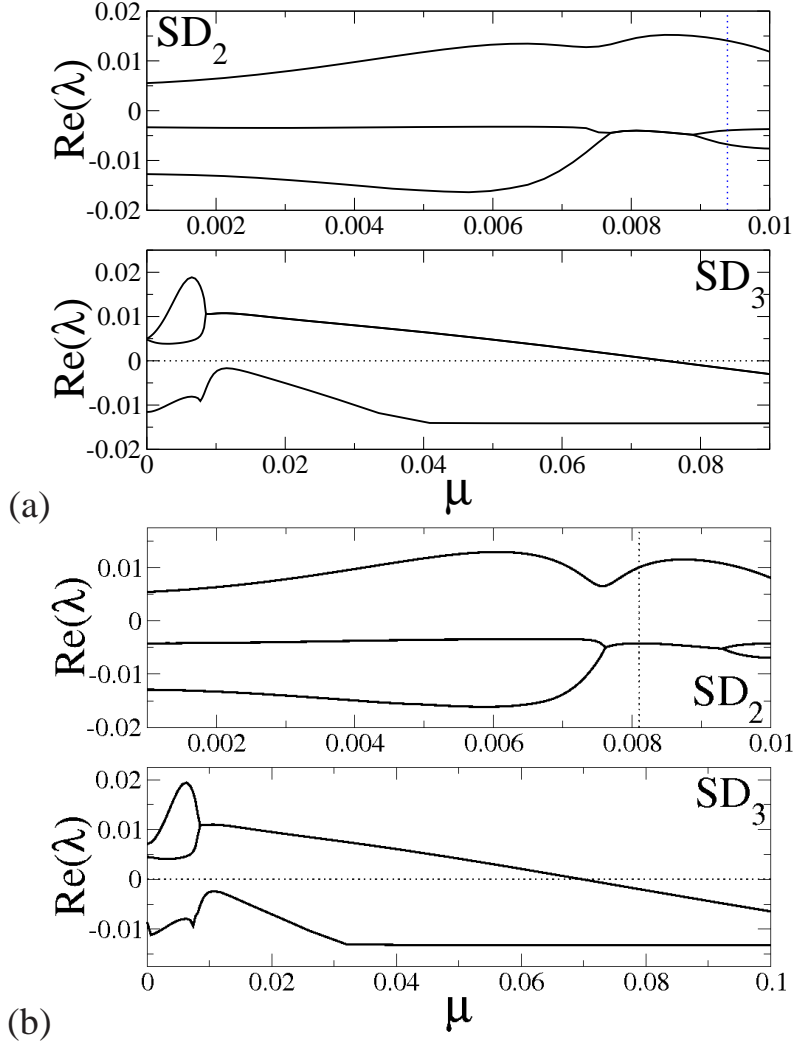}
\caption%
{Real part of the leading eigenvalues of the SD$_2$ and SD$_3$ branches for
  (a) $L=26$ (Fig.~\ref{fig:cpnel26}) and (b) $L=27$ (Fig.~\ref{fig:cpnel27}).  
  The dotted line indicates the approximate location of the global bifurcation,
  i.e., $\mu=a_1$ (Fig.~\ref{fig:period-fits-SSD}). As $\mu$ increases the 
  SD$_3$ branch stabilizes via a Hopf 
  bifurcation at $\mu_{\mathrm{hopf}}^\mathrm{3d}\approx0.075$ ($L=26$) and at
  $\mu_{\mathrm{hopf}}^\mathrm{3d}\approx0.07$ ($L=27$), respectively.  }%
\mylab{fig:evSD2ell26}
\end{figure}

\begin{figure}[h]
\centering
\includegraphics[width=0.9\hsize]{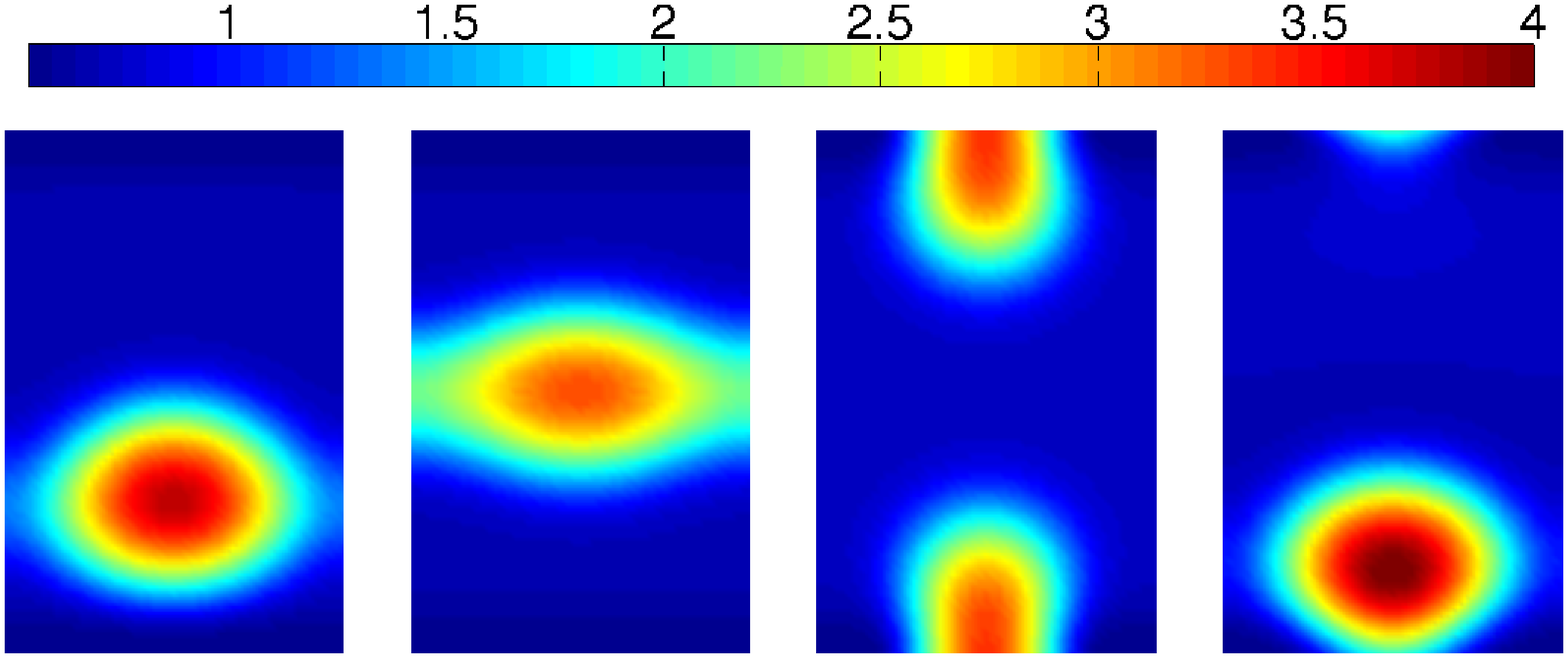}\\
{\large (a)\hfill(b)\hfill(c)\hfill (d)}
\caption%
{(color online) Scenario (ii-b): Steady states at locations indicated by open squares
in~\myfig{cpnel26} in terms of contours of constant $h(x,y)$.
Drops at (a) the saddle-node $\mu_{\mathrm{sn2}}^\mathrm{3d}\approx 0.0020$, and on the SD$_2$ 
branch at (c) $\mu=0$ and (d) $\mu=0.0070$.  The modulated ridge 
shown in (b) is also at $\mu=0$.  }
\mylab{fig:snapel26}
\end{figure}

For $L>L_\mathrm{sn1}^0$ two 3d solutions are present (in addition to
the previously studied 2d SR solutions) even at $\mu=0$. Another two
appear at $L=L_\mathrm{sn2}^0$ slightly above $L_\mathrm{sn1}^0$
(cf.~Fig.~\ref{fig:celmu0eps0,3}).  The resulting bifurcation diagram
is shown in Fig.~\ref{fig:cpnel26}.  As in scenario (ii-a), the pinned
2d ridge is linearly stable at small $\mu$ but becomes unstable
w.r.t.\ 3d perturbations at
$\mu_\mathrm{c2}^\mathrm{3d}(L)<\mu_\mathrm{hopf}^\mathrm{2d}<\mu_\mathrm{sn1}^\mathrm{2d}$.
Continuation of the branch of SD$_2$ solutions when $L=26$ shows a
branch that bifurcates subcritically from the SR$_1$ branch at
$\mu_\mathrm{c1}^\mathrm{3d}\approx 0.0111$ (marked by a cross in
Fig.~\ref{fig:cpnel26}) and continues all the way to $\mu=0$.  The two
solutions with the largest norm at $\mu=0$ correspond to stable and
unstable steady drop (SD$_1$) solutions, connected via a saddle-node
bifurcation at $\mu_\mathrm{sn2}^\mathrm{3d}\approx 0.002019$. Such
solutions are present whenever $L>L_\mathrm{sn1}^0$. Time simulations
at $\mu$ slightly above $\mu_\mathrm{sn2}^\mathrm{3d}$ with a stable
steady drop as initial condition show that the saddle-node bifurcation
is not a sniper bifurcation: the steady drop does not start to slide
but instead stretches in the $y$-direction and converges to the stable
2d SR$_1$ state.

The transition from (ii-a) to (ii-b) is now clear: with increasing
$L$ the saddle-node bifurcation between the branches SD$_2$ and SD$_3$ 
[at $\mu_\mathrm{sn1}^\mathrm{3d}$, cf.~case (ii-a)] moves towards $\mu=0$. 
At $L=L_\mathrm{sn2}^0$ it touches the line $\mu=0$ and two new 
solutions appear at $\mu=0$ corresponding to the SD$_2$ and SD$_3$ branches
in Fig.~\ref{fig:celmu0eps0,3}). In contrast, the behavior of the sliding 
solutions remains unclear. The 2d oscillations and stick-slip states 
(Fig.~\ref{fig:cpnob2d}) continue to exist but are now unstable with respect 
to 3d perturbations, and time-simulations result in stick-slip drops (SSD).  
Since this branch does not terminate in a local bifurcation we conjecture
that it terminates in a global bifurcation on the (unstable) SD$_2$
branch. The discussion in section \ref{sec:int} supports this
interpretation as does Ref.~\cite{Sigg03}.

In Fig.~\ref{fig:period-fits-SSD} we show logarithmic fits to the SSD 
period $T$, i.e., the time it takes for a drop to slide
to the next downstream defect following depinning, as a function
of $\mu$ for $L=26$ (cf.~Fig.~\ref{fig:cpnel26}) and $L=27$
(cf.~Fig.~\ref{fig:cpnel27}). The points near the critical
parameter value $\mu=a_1\approx\mu_g^\mathrm{3d}$ are given the highest 
weight. A square-root power law of the type expected near a sniper 
bifurcation is not compatible with the data while the logarithmic fit shown
in the figure provides strong evidence for depinning via a global bifurcation
for both $L=26$ and $L=27$. We caution, however, that fits of this type cannot 
exclude the presence of a fold in the SSD branch prior to depinning 
($\mu\rightarrow\mu_g^\mathrm{3d}$, $T\rightarrow\infty$) because of the absence of very 
high period simulations. On the other hand, the figure clearly differentiates 
between the global bifurcations present for $L=26$ and $L=27$ and sniper 
bifurcations present for $L=27.8$ (cf.~Fig.~\ref{fig:cpnel27,8} below) and 
larger $L$, for which a square-root fit works very well.

The logarithmic fits provide information about the leading eigenvalues of the 
saddle-type equilibria responsible for the presence of the global bifurcation
suggested by the fits. We assume that the two leading eigenvalues of the 
equilibrium at $\mu=a_1$ (i.e., the eigenvalues closest to zero) are real, with 
one positive eigenvalue $\lambda_u>0$ and one negative eigenvalue $\lambda_s<0$. 
We assume that all the remaining eigenvalues, whether real or complex, are stable 
with modulus larger than $|\lambda_s|$. Under these circumstances standard theory 
shows that near $\mu=a_1$, the period follows the asymptotic behavior 
$T\approx - \frac{1}{\lambda_u}\ln(\mu-a_1)$ provided $\lambda_u+\lambda_s<0$,
and $T\approx \frac{1}{\lambda_s}\ln(\mu-a_1)$ provided $\lambda_u+\lambda_s>0$.
In the former case the periodic orbit is stable near $\mu=a_1$; in the latter
it is unstable. However, in both cases the growth of the period is asymptotically
monotonic with the distance from $\mu=a_1$.

We have computed the leading eigenvalues along the SD$_2$ and SD$_3$
branches for the two values of $L$. Both branches are of saddle-type near
the global bifurcation and hence candidates for involvement in the
global bifurcation. Figure~\ref{fig:evSD2ell26} shows that for (a) $L=26$
the two leading eigenvalues of SD$_2$ at $\mu=a_1\approx 0.00939$ are $\lambda_u=0.0140$ 
and $\lambda_s=-0.00396$, while for (b) $L=27$ the leading unstable eigenvalue 
at $\mu=a_1\approx 0.00810$ is $\lambda_u=0.0101$ and the leading stable
eigenvalue is complex, with real part ${\rm Re}\lambda_s=-0.00420$. Thus 
the theory summarized above applies in the first case only; the complex eigenvalues
in the second case suggest an oscillatory approach to the global bifurcation of
Shil'nikov type that is not observed. In the first case the fact that
$\lambda_u+\lambda_s>0$ indicates that the appropriate prefactor in the scaling of 
the asymptotic period is $a_0=1/|\lambda_s|\approx252.5$. This value does not
compare well with the fit (a) in Fig.~\ref{fig:period-fits-SSD}. As in the 2d case
we surmise that this discrepancy is due to the fact that $\lambda_u+\lambda_s>0$ 
implies that if SD$_2$ is responsible for the global bifurcation the nearby periodic 
orbits will necessarily be {\it unstable}. Thus the branch of periodic states must 
in fact overshoot the global bifurcation, and double back, i.e., the branch
{\it must} undergo a saddle-node bifurcation and lose stability prior to the 
global bifurcation much as in the 2d case discussed in Fig.~\ref{fig:cpnob2d}.
However, in contrast to Fig.~\ref{fig:cpnob2d}, in the 3d case we have been 
unable to detect any evidence for a saddle-node of the SSD states in this 
region. Likewise in the second case the fact that 
$\lambda_u+2{\rm Re}\lambda_s>0$ implies that all long period SSD states will 
also be unstable.

Close to $\mu=\mu_g^\mathrm{3d}$, but before the asymptotic regime just described sets
in, the periodic orbits are, however, stable, and in this case the period should be 
approximated by $T\approx - \frac{1}{\lambda_u}\ln(\mu-a_1)$, i.e., we expect that in 
this regime the correct value of $a_0$ is $a_0\approx 71.4$ ($L=26$). This result is to 
be compared with the fit $a_0=88.3$ (Fig.~\ref{fig:period-fits-SSD}). Of course, because 
of the overshoot, the fitted value of $a_1$, $a_1\approx 0.00939$, is smaller than the 
true value $\mu=\mu_g^\mathrm{3d}$ and this is so in case (b) as well. Here the fit yields
$a_0\approx112.85$ ($L=27$) which again compares well with $\frac{1}{\lambda_u}=101.3$.

\subsubsection{Scenario (iii-a):  $L_c^0 \leq L \leq L_{1}$}

\begin{figure}[h]
\centering
\includegraphics[width=0.7\hsize]{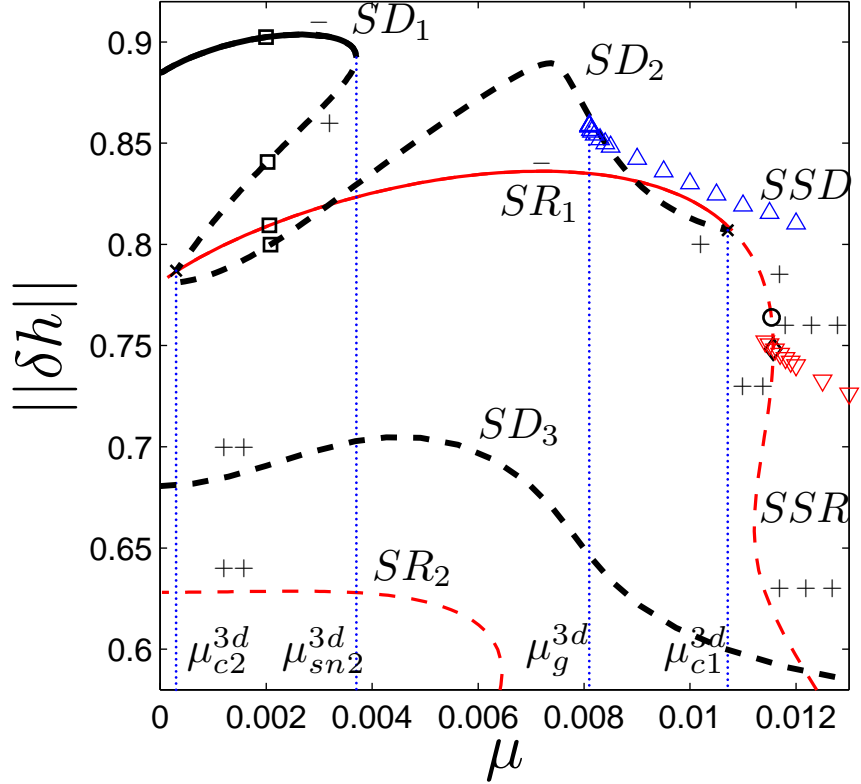}
\caption%
{(color online) Scenario (iii-a): Bifurcation diagram for steady
  ridges (SR, thin red lines), steady drops (SD, black lines) on a substrate with
  hydrophobic line defects (wettability contrast $\epsilon=0.3$) and
  spanwise system size $L=27$ showing the $L^2$ norm $||\delta h||$ of
  steady solutions as a function of the driving force $\mu$. Solid
  (dashed) lines denote stable (unstable) solutions. Downward [upward]
  pointing triangles indicate SSR [SSD] solutions.  Solution profiles
  at locations indicated by open squares are shown in
  Fig.~\ref{fig:snapel27}.  The remaining parameters are the same as
  in \myfig{celnob}.  }%
\mylab{fig:cpnel27}
\end{figure}
\begin{figure}[h]
\centering

\includegraphics[width=0.9\hsize]{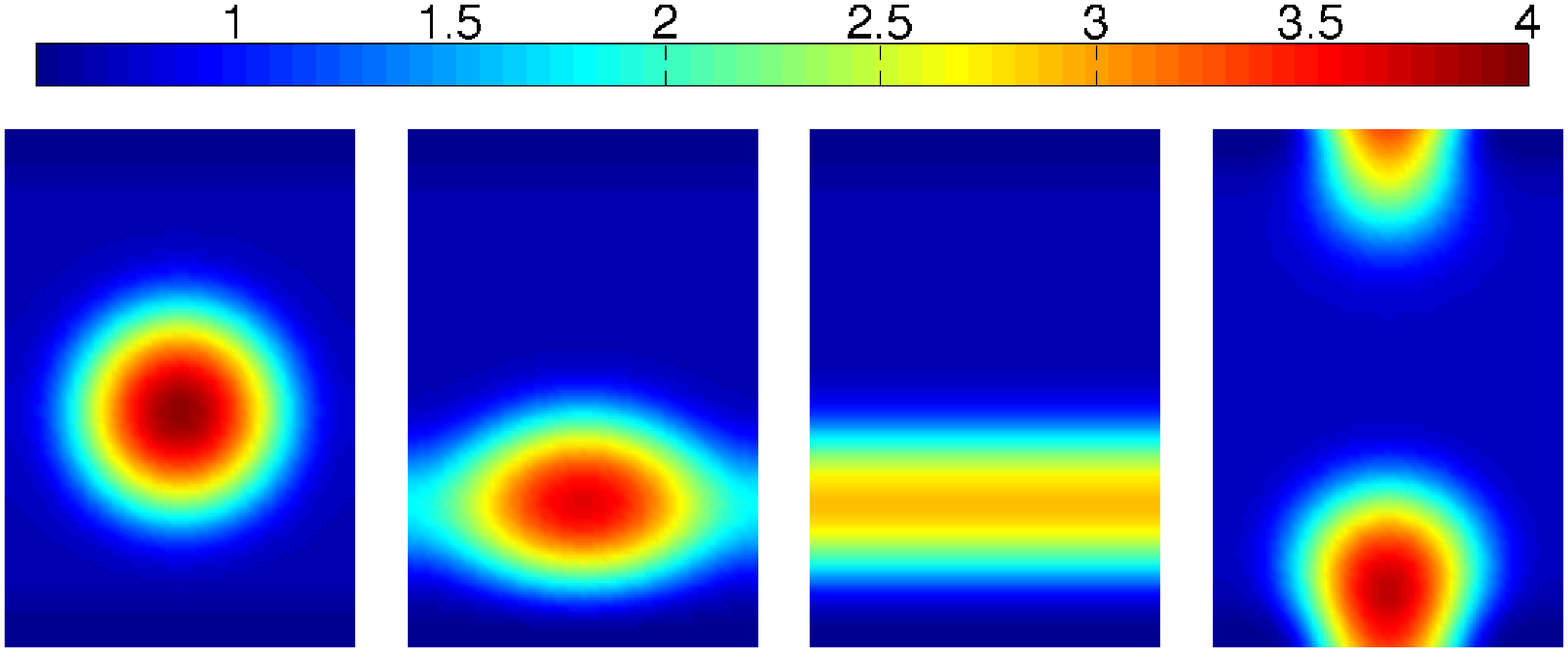}\\
{\large (a)\hfill(b)\hfill(c)\hfill(d)}
\caption%
{(color online) Scenario (iii-a): Steady states at locations indicated by open squares
in~\myfig{cpnel27} in terms of contours of constant $h(x,y)$, ordered
by decreasing norm. (a) $\mu=0.00199$, (b) $\mu=0.00203$, (c) $\mu=0.00188$ 
and (d) $\mu=0.00208$.}%
  \mylab{fig:snapel27}
\end{figure}

As $L$ increases beyond $L_c^0$ two related changes occur simultaneously at 
$\mu=0$: the 2d SR$_1$ state becomes linearly unstable and the subcritical part 
of the SD$_1$ branch is no longer present (cf.~Fig.~\ref{fig:celnob}). As a 
result when $\mu$ increases the SR$_1$ acquires stability with respect
to the Plateau-Rayleigh instability at 
$\mu_{c2}^\mathrm{3d}\approx 0.000286$ (for $L=27$) before losing it again at 
$\mu_{c1}^\mathrm{3d}\approx 0.0107783$ (Fig.~\ref{fig:cpelob}).
The resulting bifurcation diagram obtained using continuation together with 
time-stepping computations is shown in Fig.~\ref{fig:cpnel27}. Selected
steady profiles are shown in Fig.~\ref{fig:snapel27}.  

Although the linear stability properties of the SR$_1$ state do not change when
$L$ increases towards $L_\mathrm{max}$, the bifurcation diagram changes its 
appearance twice owing to codimension two bifurcations that take place at 
$L=L_{1}$ and $L=L_{2}$ (see below).  We treat the resulting $L$ regimes in 
cases (iii-a), (iii-b) and (iii-c), respectively.

The main difference between case (iii-a) and the scenario (ii-b) already 
discussed is found in the behavior of the unstable part of the SD$_1$ branch. In 
(iii-a) this branch does not extend to $\mu=0$ but terminates instead on the 
SR$_1$ branch in a subcritical bifurcation at $\mu_{c2}^\mathrm{3d}$. With increasing 
amplitude the unstable part of the SD$_1$ branch turns around in a 'true' saddle-node 
bifurcation at $\mu_\mathrm{sn2}^\mathrm{3d}\approx 0.0037015$ and 
acquires stability. Time evolution at $\mu$ slightly above the saddle-node at 
$\mu_\mathrm{sn2}^\mathrm{3d}$ starting from a stable drop at lower $\mu$ 
converges to the stable SR$_1$ solution. Examples of profiles on the SD$_1$ branch 
are shown in \myfig{snapel27}.
The bifurcation at $\mu_{c1}^\mathrm{3d}$ also generates subcritical 3d steady states. 
As in scenario (ii-b) these extend all the way to $\mu=0$, albeit with a
slightly higher and more pointed maximum norm. The norm at small
$\mu$ is also larger and is now comparable to the amplitude of the SR$_1$
states.
For $\mu>\mu_{c1}^\mathrm{3d}$ time evolution results in a family of stick-slip
drops (SSD).

\subsubsection{Scenario (iii-b): $L_{1} \leq L \leq L_{2}$}\label{sec:iii_b}

\begin{figure}[h]
\centering
\includegraphics[width=0.45\hsize]{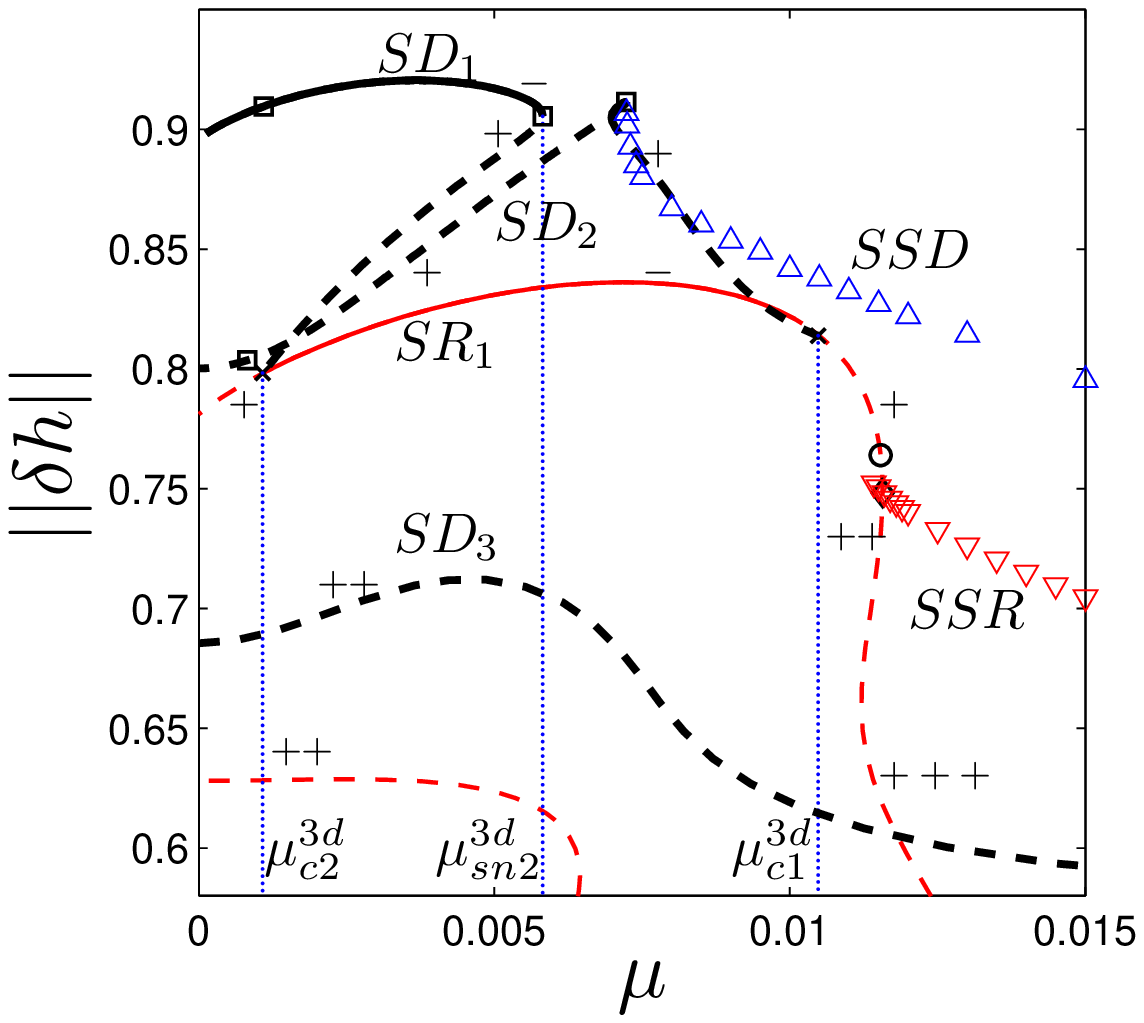}
\includegraphics[width=0.45\hsize]{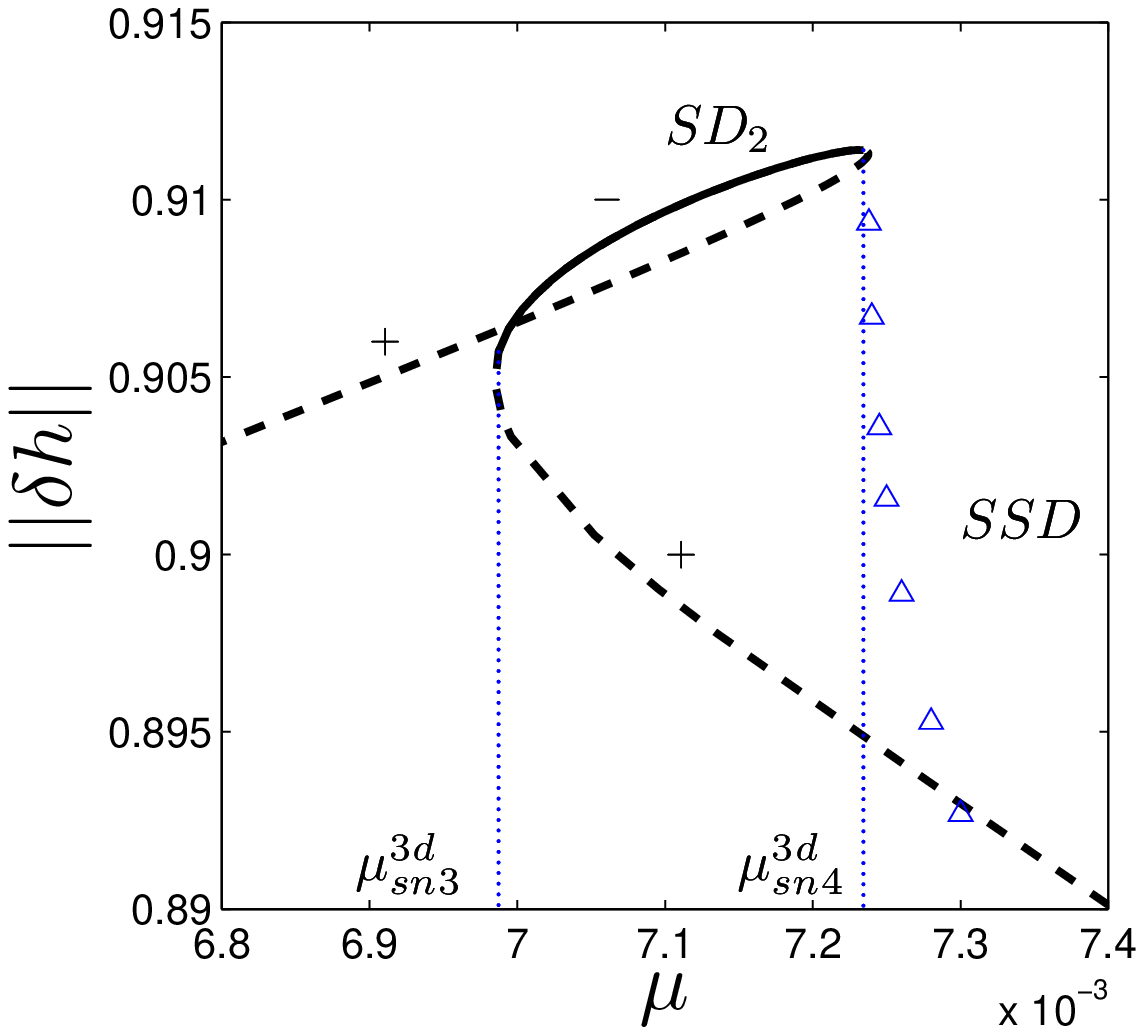}\\
{\large (a)\hfill(b)}
\caption%
{(color online) Scenario (iii-b): (a) Bifurcation diagram for steady
  ridges (SR, thin red lines), and steady drops (SD, black lines) pinned by a
  hydrophobic line defect of strength $\epsilon=0.3$ for $L=27.8$
  showing the $L^2$ norm $||\delta h||$ of steady solutions as a
  function of the driving force $\mu$. Solid (dashed) lines denote
  stable (unstable) solutions.  The panel (b) shows a zoom of the
  cusp-like feature on the SD$_2$ branch. The SD$_3$ branch becomes
  stable at $\mu=0.064$ (not shown).  Downward [upward] pointing
  triangles indicate SSR [SSD] solutions. Solution profiles at
  locations indicated by open squares are shown in
  Fig.~\ref{fig:snapel27,8}.  The remaining parameters are the same as
  in \myfig{celnob}. }
\mylab{fig:cpnel27,8}
\end{figure}

\begin{figure}[h]
\centering
\includegraphics[width=0.9\hsize]{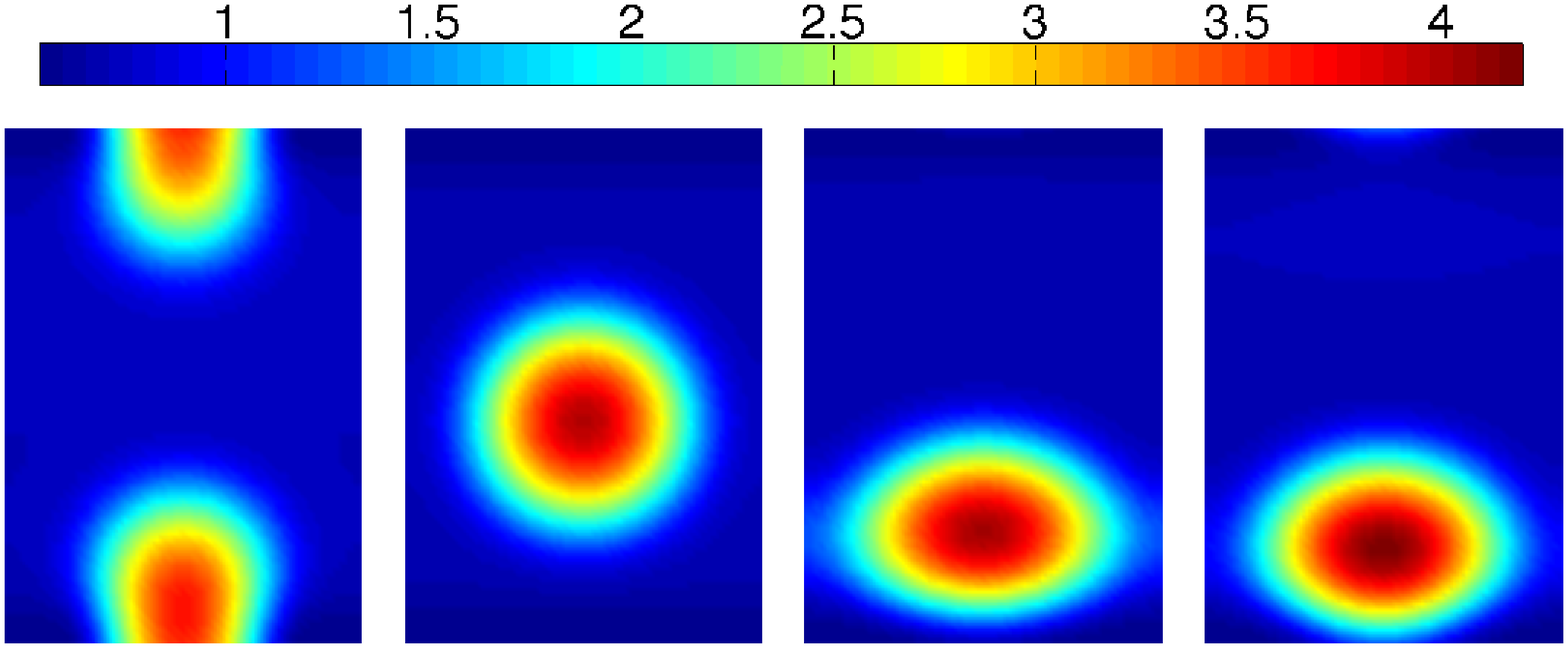}\\
{\large (a)\hfill(b)\hfill(c)\hfill(d)}
\caption%
{(color online) Scenario (iii-b): Steady states at locations indicated by open squares
in~\myfig{cpnel27,8} in terms of contours of constant $h(x,y)$.
(a) $\mu=0.00082$ on the drop branch SD$_1$, (b) on the drop branch
SD$_1$ at $\mu=0.00110$, (c)
  at the saddle-node $\mu_\mathrm{sn2}^\mathrm{3d}=0.00582$ on SD$_1$, and (d) at the 
sniper bifurcation at $\mu_\mathrm{sn4}^\mathrm{3d}=0.00724$ on SD$_2$.}
\mylab{fig:snapel27,8}
\end{figure}

At $L=L_{1}=27.59$ a codimension two bifurcation takes place and
the bifurcation diagram changes from Fig.~\ref{fig:cpnel27} to that in
Fig.~\ref{fig:cpnel27,8}(a).  The figure shows that no qualitative
change in behavior occurs at zero or small $\mu$. The SR$_1$ branch is
unstable at $\mu=0$ and stabilizes at $\mu_\mathrm{c2}^\mathrm{3d}\approx
0.0010798$ (for $L=27.8$). This bifurcation continues to be subcritical and
produces a branch SD$_1$ of unstable 3d states. The SD$_1$ states
annihilate in a 'true' saddle-node bifurcation at
$\mu_\mathrm{sn2}^\mathrm{3d}\approx 0.0058161$.  However, the 3d
solutions that emerge subcritically at $\mu_{c1}^\mathrm{3d}\approx
0.0104789$ and extend to $\mu=0$ undergo a major change near
maximum norm. At $L=L_{1}$ a cusp appears and for $L>L_{1}$ the branch
develops a loop, with two additional saddle-node bifurcations, at
$\mu_\mathrm{sn3}^\mathrm{3d}\approx 0.0069872$ and
$\mu_\mathrm{sniper}^\mathrm{3d}=\mu_\mathrm{sn4}^\mathrm{3d}\approx
0.0072342$ (for $L=27.8$), as visible in the zoom in
Fig.~\ref{fig:cpnel27,8}(b). At the same time the branch acquires a
linearly stable segment that extends between the two saddle-node
bifurcations. A sample stable solution is shown in
Fig.~\ref{fig:snapel27,8}.  Moreover, while the left-hand saddle-node
at $\mu_\mathrm{sn3}^\mathrm{3d}$ is a 'true' saddle-node, that at
$\mu_\mathrm{sniper}^\mathrm{3d}=\mu_\mathrm{sn4}^\mathrm{3d}$ now corresponds 
to a sniper bifurcation, with a branch of stick-slip drops (SSD) emerging from
$\mu_\mathrm{sniper}^\mathrm{3d}$. 

With increasing $L$ the topology of the diagram remains unchanged until a
second critical value, $L=L_{2}$, as discussed next.

\subsubsection{Scenario (iii-c):  $L_{2} \leq L \leq L_\mathrm{max}$}\label{sec:iii_c}

\begin{figure}[h]
\centering
\includegraphics[width=0.45\hsize]{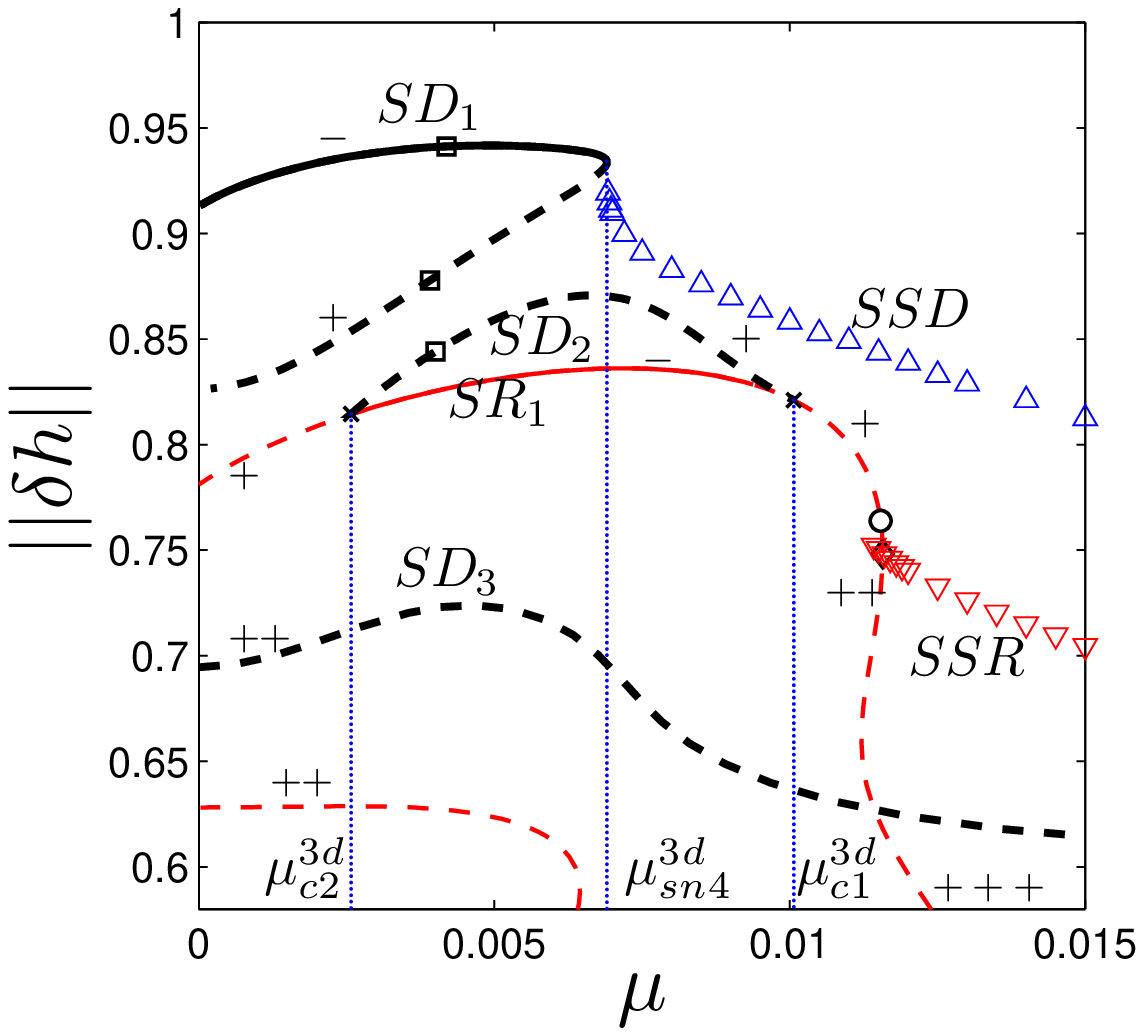}
\includegraphics[width=0.45\hsize]{fig23b}\\
{\large (a)\hfill(b)}
\caption%
{(color online) Scenario (iii-c): Bifurcation diagram for steady
  ridges (SR, thin red lines), and steady drops (SD, black lines) on a
  substrate with hydrophobic line defects (wettability contrast
  $\epsilon=0.3$) and spanwise system size $L=29$ showing the $L^2$
  norm $||\delta h||$ of steady solutions as a function of the driving
  force $\mu$. Solid (dashed) lines denote stable (unstable)
  solutions. Downward [upward] pointing triangles indicate SSR [SSD]
  solutions. Solution profiles at locations indicated by open squares
  are shown in Fig.~\ref{fig:snapel29}. The remaining parameters are
  as in \myfig{celnob}. The SD$_3$ branch becomes stable at
  $\mu=0.062$ (off scale).  (b) Real part of the leading eigenvalues
  for the stable (thin black line) and unstable part (heavy red line)
  of SD$_1$ drop branch.  The black dashed line corresponds to a
  complex eigenvalue.  }%
\mylab{fig:cpnel29}
\end{figure}

\begin{figure}[h]
\centering
\includegraphics[width=0.8\hsize]{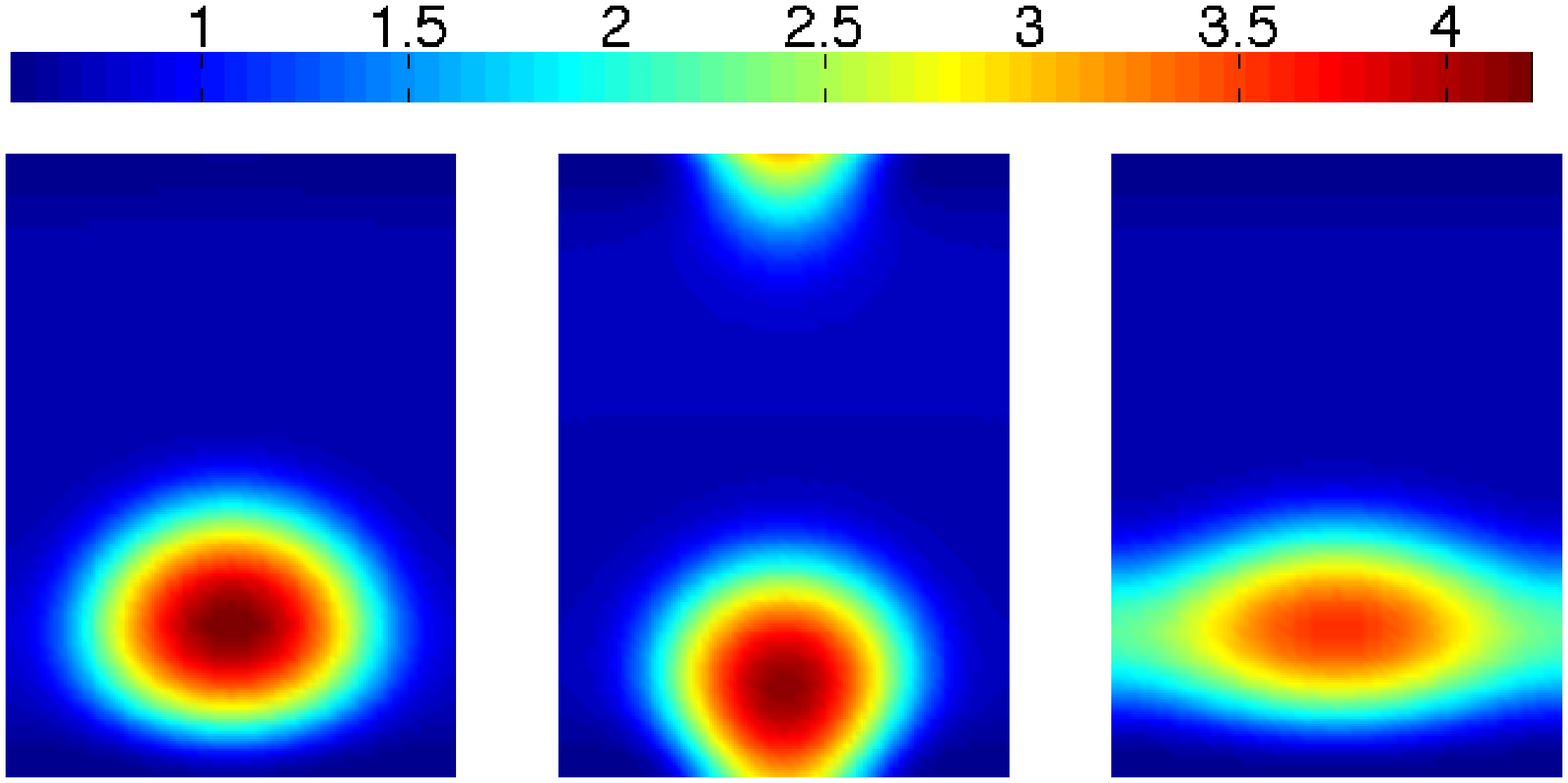}\\
{\large (a)\hfill(b)\hfill(c)}
\caption%
{(color online) Scenario (iii-c): Steady states at locations indicated by open
  squares in~\myfig{cpnel29} in terms of contours of constant
  $h(x,y)$, ordered by decreasing norm. (a) $\mu=0.00419$, (b) $\mu=0.00391$ 
  and (c) $\mu=0.00401$. The remaining parameters are as in \myfig{celnob}.  }%
\mylab{fig:snapel29}
\end{figure}

Increasing $L$ beyond $L_{2}= 27,87$ another
codimension two bifurcation takes place and the bifurcation diagram
changes from Fig.~\ref{fig:cpnel27,8}(a) to that in
Fig.~\ref{fig:cpnel29}. Once again the branches at $\mu=0$ and their
stability properties remain unchanged, as do the properties of the
SR$_1$ states. However, the nonlinear behavior at finite $\mu$ changes
dramatically with the 3d branches in Fig.~\ref{fig:cpnel27,8}(a)
reconnecting in a new way. This reconnection occurs as a result of the
fusion/annihilation of the saddle-node at $\mu_\mathrm{sn2}^\mathrm{3d}$ on
the SD$_1$ branch and the newly created saddle-node at
$\mu_\mathrm{sn3}^\mathrm{3d}$ on the branch of 3d states SD$_2$ via a ``necking''
bifurcation. As a result the branches reorder into a simpler
bifurcation pattern [Fig.~\ref{fig:cpnel29}(a)]. The stable part of
the SD$_1$ branch now annihilates with the left part of the unstable
SD$_2$ branch in a saddle-node bifurcation at
$\mu_\mathrm{sniper}^\mathrm{3d}=\mu_\mathrm{sn4}^\mathrm{3d}\approx
0.0069055$ that is still a sniper bifurcation [in
Fig.~\ref{fig:cpnel29}(a)]. As a result of the various transitions
occurring with increasing $L$ the depinning transition has shifted
from the 2d SR$_1$ states \cite{ThKn06} to the 3d SD$_1$ states \cite{BHT09},
i.e., the behavior expected of large drops in large domains. 

Figure~\ref{fig:cpnel29}(b) shows the eigenvalues along the stable and
unstable parts of the SD$_1$ branch and reveals that the leading eigenvalue 
of the stable drop solution is complex close to the saddle-node at 
$\mu_\mathrm{sniper}^\mathrm{3d}$. In this regime the stable SD$_1$ states
relax in an oscillatory fashion when perturbed. In contrast, the leading unstable 
eigenvalue along the unstable SD$_1$ branch is real, implying monotonic growth.
This eigenvalue is created by a collision of the complex eigenvalues
on the negative real axis prior to the saddle-node bifurcation creating a pair
of real eigenvalues, one of which crosses into the positive half-plane at 
$\mu_\mathrm{sniper}^\mathrm{3d}$.

At the same time the branch of unstable 3d states that emerges at
$\mu_{c2}^\mathrm{3d}\approx 0.0025782$ continues towards and ends at
$\mu_{c1}^\mathrm{3d}\approx 0.010065$. As $L$ is increased further,
$\mu_{c1}^\mathrm{3d}$ and $\mu_{c2}^\mathrm{3d}$ approach each other
until they annihilate at $L_\mathrm{max}$
(Fig.~\ref{fig:cpelob}). Consequently the branch SD$_2$ of
unstable 3d states becomes shorter and shorter, along with the
interval of stable SR$_1$ states, and when it vanishes so does the
SR$_1$ stability interval. However, for the parameter values of
Fig.~\ref{fig:cpnel29}(a) we find coexistence between stable SR$_1$
states and either stable SD$_1$ states or, beyond the sniper
bifurcation at $\mu_\mathrm{sniper}^\mathrm{3d}=\mu_\mathrm{sn4}^\mathrm{3d}$ 
(numerically stable) stick-slip drop motion (SSD).

\subsubsection{Scenario (iv):  $L_\mathrm{max} \leq L $}

\begin{figure}[h]
\centering
\includegraphics[width=0.7\hsize]{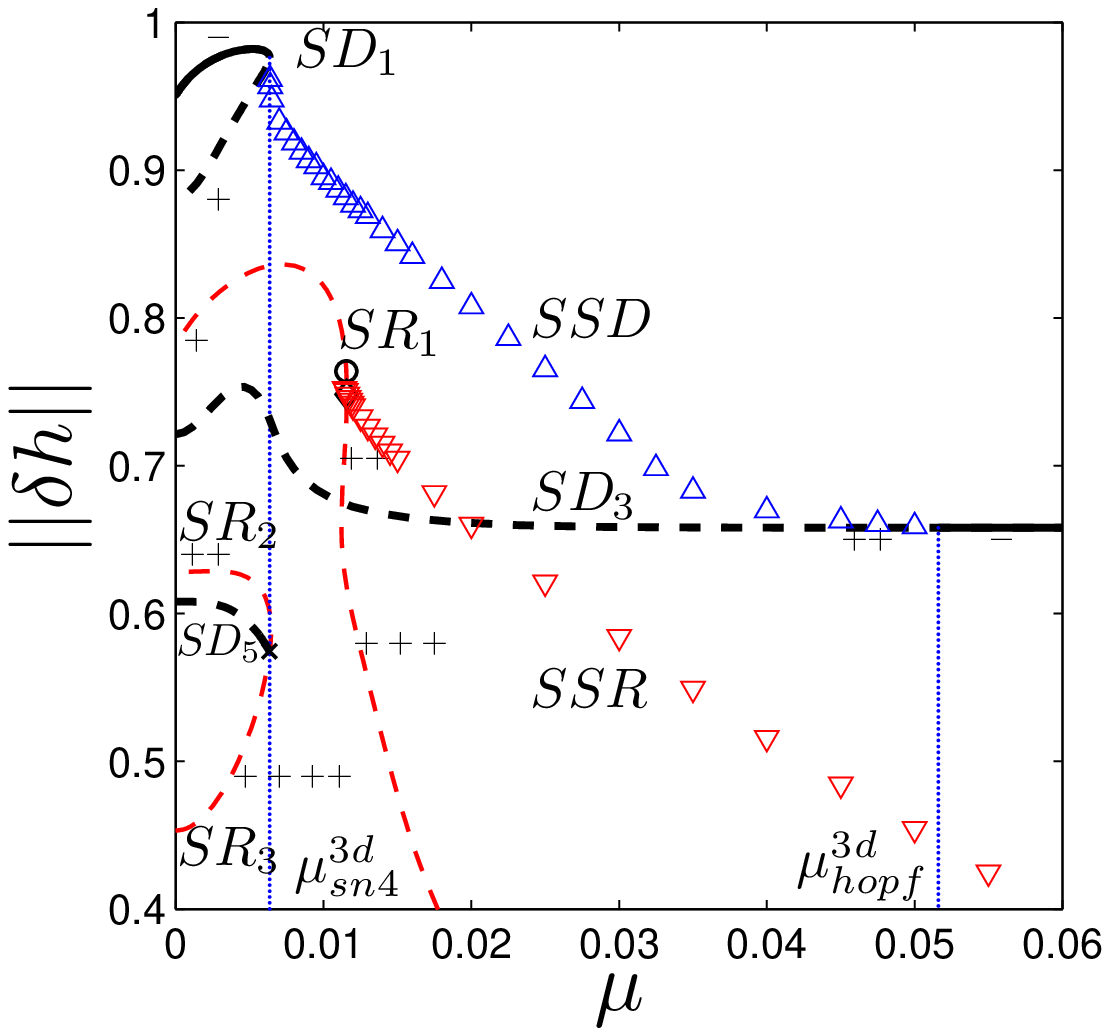}
\caption%
{(color online) Scenario (iv): Bifurcation diagram for steady ridges
  (SR, thin red lines), and steady drops (SD, black lines) on a substrate
  with hydrophobic line defects (wettability contrast $\epsilon=0.3$)
  and spanwise system size $L=32$ showing the $L^2$ norm $||\delta
  h||$ of steady solutions as a function of the driving force
  $\mu$. Solid (dashed) lines denote stable (unstable)
  solutions. Downward [upward] pointing triangles indicate SSR [SSD]
  solutions.  The remaining parameters are as in \myfig{celnob}. The
  SD$_3$ branch becomes stable at
  $\mu_\mathrm{hopf}^\mathrm{3d}\approx 0.051$.  }%
\mylab{fig:cpnel32}
\end{figure}

When $L$ increases beyond $L_\mathrm{max}$ the linear stability
problem indicates (Fig.~\ref{fig:cpelob}) that the spanwise-invariant
ridge SR$_1$ is linearly unstable for all $\mu$ with a real positive
leading eigenvalue. Thus the two pitchforks responsible for the
unstable modulated ridges annihilate as $L$ increases and the
modulated ridge states disappear (see Fig.~\ref{fig:cpnel32}). As a
result the 3d drop states SD$_1$ now correspond to the drops studied
for large domains in \cite{BHT09}. At these values of $L$ a stable and
an unstable drop of this type annihilate in a sniper bifurcation at
$\mu_\mathrm{sn4}^\mathrm{3d}\approx 0.0069$.  The emerging branch of
stick-slipping drops (SSD) gradually approaches the branch of SD$_3$
states, connecting to it in a Hopf bifurcation at
$\mu_\mathrm{hopf}^\mathrm{3d}\approx 0.051$. Since at large $\mu$ the SD$_3$
states resemble modulated rivulets [cf.~Fig.~\ref{fig:snapel23}~(c)]
close to the Hopf bifurcation the SSD states correspond to surface
waves on a steady rivulet.

\subsection{Stability in the $(L,\mu)$ plane}
\mylab{sec:all-stab}
\begin{figure}[tbh]
\centering
\includegraphics[width=0.7\hsize]{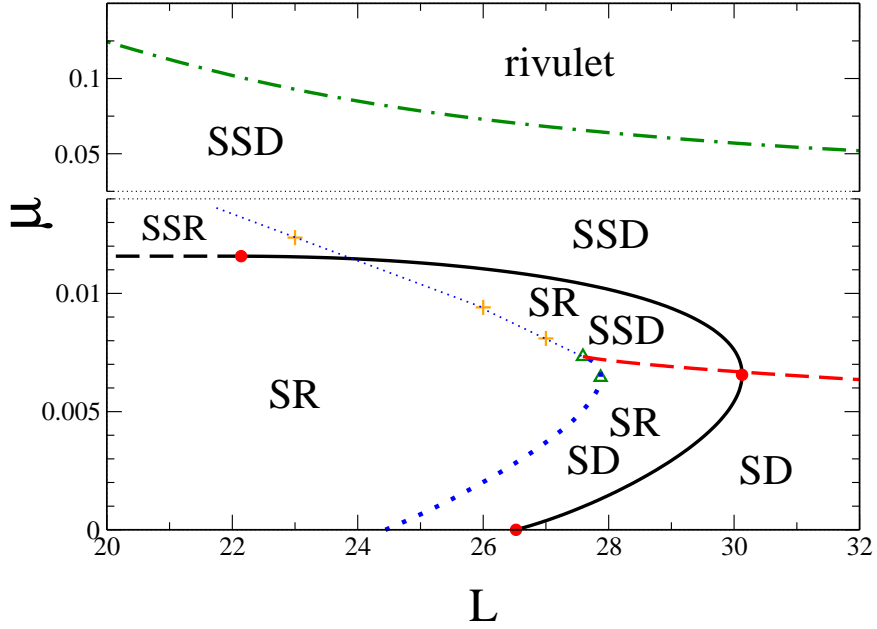}
\caption%
{(color online) Stability diagram for steady (SR) and stick-slipping
  ridge (SSR), steady (SD) and stick-slipping drop (SSD) and rivulet
  states on a heterogeneous incline showing the region of stability in
  the $(L,\mu)$ plane for $\epsilon=0.3$, ${\bar h}=1.2$, and
  $L_x=40$. Black solid (dashed) lines indicate the transversal
  (sniper) instability of the steady ridges as in \myfig{cpelob}. The
  thick dotted blue (dashed red) lines are the loci of saddle-node
  (sniper) bifurcations of the drop states. The thin dotted blue line
  indicates the hypothetical border of the region of stable
  stick-slipping drops (the three ``+'' symbols result from our
  calculations). Finally, the dot-dashed green line indicates the loci
  of the Hopf bifurcation where the steady 3d rivulets become stable
  when $\mu$ increases.}%
\mylab{fig:diag_stab_ss}
\end{figure}
The bifurcation analysis in the preceding sections
  has identified several types of stable steady (ridge, drop and
  rivulet) and time-periodic (stick-slipping ridge and stick-slipping
  drop) states. Figure~\ref{fig:diag_stab_ss} displays the stability
  regions of these states in the $(L,\mu)$ parameter plane. Most of
  the boundaries of these regions are computed by numerically tracking
  the bifurcations that lead to loss of stability. For example, since
  the drop state loses stability via a saddle-node or a sniper
  bifurcation, one of the stability boundaries requires the tracking
  of a zero eigenvalue of the linear problem obtained by linearizing
  Eq.~(\ref{eq:lub}) about the drop state.  This is accomplished using
  a steady-state continuation algorithm similar to that described in
  \cite{BeTh10}.  Likewise, since the rivulet state is stabilized via
  a Hopf bifurcation numerical tracking of the Hopf bifurcation yields
  the stability boundary of the rivulet state.

The region of stable steady ridge solutions is discussed in
  \mysec{incl2s}.  Note that it has some overlap with the region of
  stable steady drop states. This indicates bistability and implies
  that, in the intersection region the observed state depends on
  initial conditions. The region of stable steady drop solutions is
  delimited by two curves which end in a cusp (upper green triangle)
  corresponding to the codimension two bifurcation at $(L_1,\mu_1)$
  (see \mysec{iii_b}). The leftmost boundary of the SD region (thick
  blue dotted line in \myfig{diag_stab_ss}) starts from the point
  $(L^0_{sn1},\mu=0)$ and terminates at the cusp. The turning point at
  $(L_2,\mu_2)$ (lower green triangle) corresponds to the codimension
  two bifurcation identified in \mysec{iii_c}. This dotted boundary is
  the locus of 'true' saddle-nodes, i.e., a drop destabilized at this
  boundary does not start to slide but instead stretches and converges
  to a stable ridge solution (see \mysec{ii_b}). In contrast, the
  upper boundary of the SD region (red dashed line in
  \myfig{diag_stab_ss}) corresponds to sniper bifurcations, and loss
  of stability along this boundary results in periodic stick-slip
  motion. This scenario occurs even if the sniper boundary intersects
  the stable ridge region. The resulting stick-slip motion is stable
  as shown by time integration, and corroborated in the next
  section.

We are not, however, able to compute in detail the region of
  stable stick-slip motions for $L<L_1$ since this requires the
  knowledge of the loci of global homoclinic bifurcations, a difficult
  task for a problem of this complexity. Instead we indicate the
  hypothetical boundary of the region of stick-slip drop motion by a
  thin blue dotted line, guided by the orange crosses. These indicate
  the 'end' of the branch of stick-slipping drops as found in time
  simulations for selected $L$. Note, however, that the transition at the thin
  blue dotted line involves some hysteresis, at least outside the
  steady-ridge region, where simulations indicate some overlap between regions of
  stable stick-slip ridges and stick-slip drops. The extent of this
  hysteresis region remains unknown.

At large $\mu$ one encounters a region of stable rivulets.
This region does not intersect the other stability regions (note the 'broken'
$y$-axis in \myfig{diag_stab_ss}).  The critical driving
$\mu_{hopf}(L)$ corresponding to the stability boundary increases rapidly with decreasing $L$.
%
\section{Interpretation}
\mylab{sec:int}
In this section we provide a bifurcation theory interpretation of the
transition sequences described in the preceding section. We focus on
the transitions observed for $L=29$, $L=27$ and $L=23$ as the forcing
$\mu$ varies. These transitions involve first and foremost the
saddle-node bifurcation(s) on the steady ridge branch SR$_1$
(hereafter SR). In addition to this bifurcation we have identified two
nearby bifurcations, a Hopf bifurcation that leads to time-periodic
oscillations of the ridge state, and a steady state bifurcation that
breaks the invariance of the ridge state in the spanwise direction,
leading to steady drop-like states (SD).  In addition, we have found
two time-dependent solution types, sliding stick-slip ridges (SSR) and
sliding stick-slip drop-like states (SSD).

Our computations reveal that both the Hopf and symmetry-breaking steady 
state bifurcations occur close to the upper saddle-node bifurcation,
suggesting that the different bifurcation scenarios may be
understood by examining the interaction between these three bifurcations,
i.e., a codimension three bifurcation. The inclusion of the nearby lower 
saddle-node bifurcation would require a study of codimension four in
which a hysteresis bifurcation interacts with both a Hopf bifurcation
and a steady-state bifurcation. Bifurcations of this complexity have not 
been studied in full detail, eg.~\cite{HaLa07}. Moreover, the standard
approach using normal form theory describes solutions in R$^3$, the dimension
of the (extended) center manifold for a mode interaction of this type. This
reduction in dimension is a consequence of normal form symmetry which 
decouples the phase associated with the Hopf bifurcation from its amplitude.

The three-dimensional phase space R$^3$ allows of course complex (chaotic) 
dynamics \cite{ACST85,ARS04} but is unable to accomodate the sniper bifurcations that
lead to sliding states. For this purpose the problem must be posed in
R${^2}\times$S$^1$, i.e., with periodic boundary conditions inherited from
the periodicity of the problem in the downstream direction. This type of phase 
space allows two types of periodic orbits that oscillate about an
equilibrium (librations) and those for which the periodic variable increases
by $2\pi$ each period (rotations). These solution types can be visualized
in terms of the standard pendulum, with librations corresponding to 
standard pendulum oscillations and rotations corresponding to a whirling
pendulum.

Although no rigorous normal form theory exists for this case (the rotations
rely on global properties of the phase space) model problems with the
required properties are readily constructed (eg.~\cite{ARS04}). The most 
relevant such model was constructed by Krauskopf and Oldeman \cite{KrOl04} 
who investigated the properties of the following system of equations:
\begin{eqnarray}
\dot{\rho}&=&\nu_1 \rho - a \rho\sin\varphi - \rho^3,\label{nf1}\\
\dot{\varphi}&=&\nu_2+s\rho^2+2\cos\varphi+c\rho^4.\label{nf2}
\end{eqnarray}
The model can be viewed in two ways, as an interaction between a saddle-node 
bifurcation in a periodic orbit (sniper) and a Hopf bifurcation, or an
interaction between a sniper and a symmetry-breaking pitchfork bifurcation.
In the former the variable $\rho>0$ corresponds to the amplitude of the periodic
oscillations (the temporal phase decoupling in ``normal form'') while in the
latter $\rho$ corresponds to the amplitude of the state created in the 
symmetry-breaking pitchfork bifurcation. Both interpretations rely on the
symmetry $\rho\rightarrow -\rho$ of the model and are relevant to the depinning
problem.

We examine first the depinning problem in which the steady ridge state (SR)
does not undergo a Hopf bifurcation. In this case the system (\ref{nf1})-(\ref{nf2}) 
describes the dynamics near the codimension two point $(\mu_{CT},L_{CT})$ at which the 
symmetry-breaking pitchfork of the SR state coincides with the saddle-node bifurcation
at $\mu=\mu^{\rm 2d}_{\rm sn1}$. Thus $(\nu_1,\nu_2)$ represent linear combinations of
$\mu-\mu_{CT}$ and $L-L_{CT}$. In this case the variable $\rho>0$ represents the 
amplitude of the translation invariance breaking instability creating the SD
state, the spanwise phase of the mode having decoupled because of 
translation invariance of the SR state in the spanwise direction.  
The variable $\varphi$ plays the role of the downstream coordinate. Thus
equilibria with $\rho=0$ represent SR states while periodic solutions with
$\rho=0$ correspond to sliding ridges (SSR). Equilibria with $\rho\ne0$ correspond to
stationary drops (SD) while periodic solutions with $\rho\ne0$ are of two types,
oscillations about an SD state (librations) and sliding drops (rotations, SSD).
\begin{figure}[h]
\centering
\includegraphics[width=0.8\hsize]{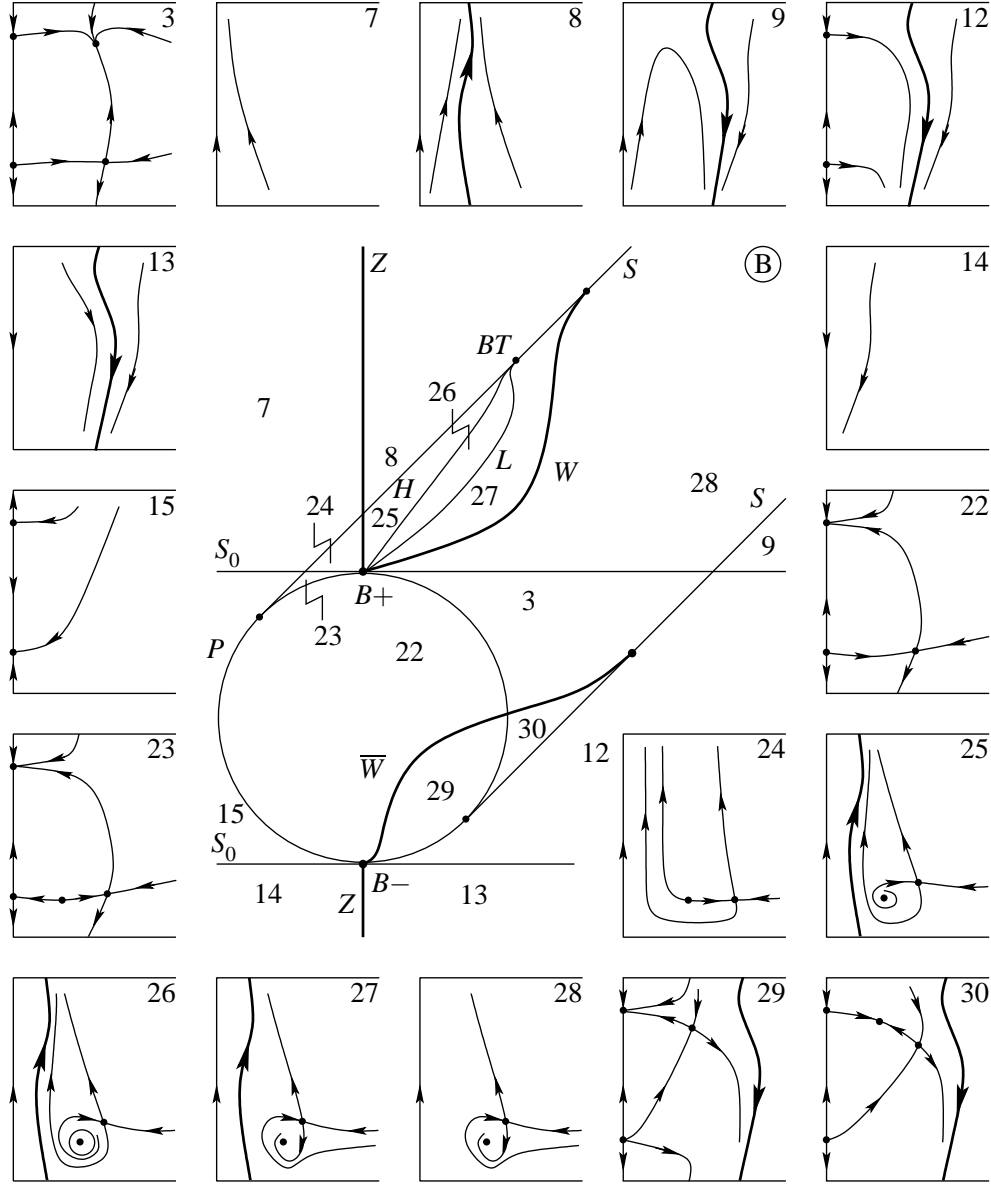}
\caption%
{The $(\nu_1,\nu_2)$ parameter plane for the system (\ref{nf1})-(\ref{nf2}) with 
$a=0.5$, $c=0$, $s=-1$ (case B of \cite{KrOl04}) showing the codimension one 
lines $S_0$, $P$, $S$ and $Z$ discussed in the text. The Takens-Bogdanov point is
labeled BT. The remaining curves correspond to global bifurcations. The phase 
portraits around the periphery show solutions characteristic of each of the 
numbered regions in the $(\rho,\phi)$ plane. Reproduced with permission from
\cite{KrOl04}.}
\mylab{fig:KrOl04fig7}
\end{figure}

In Fig.~\ref{fig:KrOl04fig7} we reproduce Figure~7 of
\cite{KrOl04}. The central part of the figure shows the
$(\nu_1,\nu_2)$ parameter plane; in the depinning problem these
parameters correspond to different combinations of the parameters
$\mu$ and $L$. The $(\nu_1,\nu_2)$ plane is split into a number of
distinct regions, labeled by integers, with distinct phase space
dynamics. The phase portraits characteristic of each region are shown
along the periphery of the diagram, with the variable $x$ shown
horizontally and $\varphi$ vertically. The lines $S_0$ in the $(\nu_1,\nu_2)$ plane
correspond to saddle-node bifurcations with $\rho=0$, i.e., to snipers on the SR
state, and are located at $\nu_2 =\pm 2$. The ellipse $P$ corresponds
to pitchfork bifurcations of equilibria with $\rho=0$, i.e., to steady
bifurcations from SR to SD. The line $S$ corresponds to saddle-nodes
of SD states (the point labeled BT is a Takens-Bogdanov point with a
double zero eigenvalue). Finally the curves labeled $Z$ indicate the
transition from SSR states to SSD states, i.e., transitions from a rotation
with $\rho=0$ to a rotation with $\rho>0$. The curves $S_0$, $P$ and $Z$
meet at points labeled B$^{\pm}$. The additional lines all
correspond to different types of global bifurcations detailed in
\cite{KrOl04}.

We are now in a position to interpret the dynamics observed in this paper in
terms of the truncated ``normal form'' (\ref{nf1})-(\ref{nf2}). Specifically, the
pinning transition observed as $\mu$ decreases, for example, for $L=32$ 
(Fig.~\ref{fig:cpnel32}), corresponds to traversing from region 13 (unstable 
SSR, stable SSD) to region 12 (a pair of unstable SR and a stable SSD) to 
region 3 (a pair of unstable SR and a pair of SD, one stable, one unstable). 
The transition from region 12 to region 3 is thus a standard sniper, except 
that the equilibria created in the transition are of SD type.

When $L=27$ the depinning transition summarized in Fig.~\ref{fig:cpnel27} is 
consistent with a slightly different cut through the $(\nu_1,\nu_2)$ plane. 
This time we pass from region 13 to region 12 followed by regions 29 (a pair 
of SR states, one of which is stable and the other unstable, an unstable SD 
state created in a subcritical pitchfork bifurcation, and a stable SSD state) 
and region 22 (a pair of SR states, one of which is stable and the other 
unstable, and an unstable SD state). Note that the disappearance of the SSD 
state is the result of a standard homoclinic bifurcation, a collision of a 
periodic orbit with a single saddle-type equilibrium, here an SD state, as 
suggested by Fig.~\ref{fig:period-fits-SSD}. Ref.~\cite{KrOl04} shows that 
these transitions are robust for $a>0$, $c=0$, $s<0$ [Eqs.~(\ref{nf1})-(\ref{nf2})]
and represent the simplest transitions from among the possibilities
shown in Fig.~\ref{fig:KrOl04fig7} and related figures valid in other
parameter regimes.

We next consider the transition observed for $L=23$
(Fig.~\ref{fig:cpnel23}).  We have not been able to interpret this
transition within the two-dimensional system
(\ref{nf1})-(\ref{nf2}). The main obstacle for this interpretation is
the coexistence of a stable SSR state with a pair of {\it unstable} SR
states, a possibility that is not permitted within Eqs.
(\ref{nf1})-(\ref{nf2}) when $\rho=0$. Indeed, Fig.~\ref{fig:cpnel23}
indicates that in this case the SR state depins via a Hopf
bifurcation, a possibility that can only be captured by a higher
codimension normal form. On the other hand the observed transition
from SSD to SSR resembles that along the line $Z$ between regions 13 
and 14 in Fig.~\ref{fig:KrOl04fig7}, except for the fact that in 
Fig.~\ref{fig:cpnel23}(b) this transition is apparently subcritical.

Since the ``normal form'' (\ref{nf1})-(\ref{nf2}) was constructed to describe the 
interaction between a Hopf and a sniper bifurcation we can also examine its
applicability in the two-dimensional depinning problem, i.e., for spanwise 
invariant ridge states. Figure~\ref{fig:cpnob2d}(b) shows the transition 
observed when three-dimensional instabilities are suppressed. The same
transition forms part of the behavior in Fig.~\ref{fig:cpnel23} despite
the presence of 3d instability. To use Fig.~\ref{fig:KrOl04fig7} we need to 
recall that fixed points of Eqs.~(\ref{nf1})-(\ref{nf2}) with $\rho>0$ now correspond 
to constant amplitude {\it oscillations} (OR) about an SR state. Thus the
standard sniper bifurcation corresponds to a transition from region 14 to region 
15. But when a Hopf bifurcation occurs on SR prior to the
saddle-node at $\mu^{\rm 2d}_{\rm sn1}$ the sniper bifurcation is replaced 
by a new depinning transition as described in \S\ref{sec:incl-ridge}.
The SSR state now involves oscillations about the SR state and so
is represented in Fig.~\ref{fig:KrOl04fig7} by a rotation at finite $\rho$.
As seen from the behavior of the period in the inset in Fig.~\ref{fig:cpnob2d}(b) 
this state disappears in a global bifurcation. We expect that this global
bifurcation involves the unstable oscillations (librations) created at
$\mu^{\rm 2d}_{\rm hopf}$. In Fig.~\ref{fig:KrOl04fig7} this sequence of transitions
is described by traversing from region 13 (stable SSR state with superposed
oscillations) into region 12 (a pair of unstable SR states together with a 
stable SSR state with superposed oscillations) and then into region 29
(a pair of SR states, one of which is stable and the other unstable, together
with an unstable libration and a stable SSR state with superposed oscillations).
The pinning transition takes place on crossing from region 29 into region 22
(a pair of SR states, one of which is stable and the other unstable, together
with an unstable libration) and so involves the formation of a homoclinic
connection to a periodic orbit. Note that in this scenario the SSR state in 
Fig.~\ref{fig:cpnob2d}(b) is a {\it two-frequency} state while the libration 
that is left has a single {\it finite} frequency. 

Our calculations indicate that the SSR state is in fact a single-frequency 
state. As a result we favor a different explanation for the behavior shown in 
Fig.~\ref{fig:cpnob2d}(b). This behavior involves a global bifurcation between 
a single frequency rotation (SSR) and a single-frequency oscillation (OR) about 
an SR state, a transition that is organized by a Takens-Bogdanov bifurcation 
with periodic reinjection. At the Taken-Bogdanov point the Hopf bifurcation 
coincides with the saddle-node and the Hopf frequency {\it vanishes}. Consequently
the ``normal form'' (\ref{nf1})-(\ref{nf2}) no longer applies. The Takens-Bogdanov point 
can be identified by shifting the parameter $\epsilon$ appropriately and 
recomputing the leading eigenvalues along the SR branch to find the point 
$(\mu_{TB},\epsilon_{TB})$ at which the zero eigenvalue has double multiplicity.
Although we have not performed such a calculation we expect this point to 
lie close to our chosen parameter values, implying that this bifurcation is 
likely relevant to the observed dynamics. The normal form for this bifurcation
is also two-dimensional and we conjecture that in the presence of reinjection
it exhibits the necessary gluing bifurcation. Similar bifurcations are known to
occur in spontaneous parity breaking in inhomogeneous systems \cite{DHK97}.
However, a study of this problem along the lines of \cite{KrOl04} is beyond the
scope of this paper.
 
The virtue of relating our results to something like normal form analysis is
that the theory is able to identify ``all'' types of dynamical behavior that
may be present in the vicinity of the origin in the $(\nu_1,\nu_2)$ plane 
[resp., $(\mu-\mu_{TB},\epsilon-\epsilon_{TB})$], thereby 
indicating the {\it possible} dynamics that may be present in the depinning 
problem for other parameter values. On the other hand the restriction to a 
low-dimensional center manifold precludes the presence of complex dynamics.
Such behavior is typically associated with the presence of global bifurcations
such as those taking place along the curves $L$, $W$ and $\overline W$ in 
Fig.~\ref{fig:KrOl04fig7}. In addition, the theory cannot account for bifurcations
that occur ``far'' from the codimension-two point. Thus one does not expect 
to be able to explain all aspects of the computed bifurcation diagrams using
low codimension normal form theory, and this is indeed the case.

%
\section{Conclusions}
\mylab{sec:conc}

In this paper we have studied the process of depinning of driven
liquid drops and ridges on heterogeneous substrates and related 
transitions between drop and ridge states. In the absence of driving, 
e.g., on a horizontal substrate, the latter transition is related to 
the Plateau-Rayleigh instability of a liquid ridge.  To explore the 
three-dimensional dynamics in the presence of driving, we have 
formulated the depinning process as a bifurcation problem, and focused 
on a generic problem of this type.

We adopted a simple model problem in which the spatial heterogeneity
corresponds to a modulation of the short range polar contribution to
the disjoining pressure and takes the form of parallel hydrophobic
stripes with a well-defined spatial period. The stripes are oriented
transverse to the driving direction, and may block the passage of 
ridges and drops, producing strongly asymmetric ridges prior to
depinning. In general, pinning might arise from spatially varying 
wetting properties as here, or from heterogeneous topography, temperature or 
electric fields. Possible driving forces include gravitational or 
centrifugal forces, and/or gradients of wettability, temperature or 
electric fields. In the present work we have used gravity as an
example, i.e., we have studied drops and ridges on an incline. As a
result our approach is limited to drops not much thicker than the 
wetting layer assumed to be present in our formulation.

The two-dimensional version of this problem was formulated and studied
in Ref.~\cite{ThKn06}, where two possible depinning mechanisms were
identified: depinning via a sniper bifurcation and depinning via a
Hopf bifurcation followed by a global bifurcation that is required for
translation of the ridge from one defect to the next. In the present
paper we have been interested in like behavior for fully three-dimensional
states that we variously refer to as modulated ridges or drops
depending on their appearance. Preliminary work on the 3d problem
\cite{BHT09} using the numerical procedure described in \cite{BeTh10}
indicated that for relatively large drops the 3d problem behaves
qualitatively like the 2d problem but the detailed depinning mechanism
was not studied owing to the long time scales involved in the
transition region. In this paper we have therefore focused on the
details of the depinning transition in the 3d case.

In three dimensions a transverse liquid ridge has additional avenues
open to it since the whole ridge no longer needs to depin
simultaneously but can instead send out protruberances through which
the fluid flows over the heterogeneity. In this paper we have
identified seven distinct regimes depending on the spanwise spatial
period $L$ of the computational domain. This spanwise length, if
sufficiently small, can suppress vestiges of the Rayleigh-Plateau
instability that leads to spanwise modulation of a ridge even on a
horizontal substrate. However, on an inclined substrate the choice of $L$
may not only allow a 3d instability to proceed but it also determines
the volume of liquid that must flow across the heterogeneity and
hence the flux of liquid through the rivulets that form as a result of
depinning. Thus when $L$ is large three-dimensional instabilities are
always present and the depinning of a ridge is preceded by an
instability of the 2d ridge that forms 3d drop-like states that remain
pinned to the heterogeneity, with a further increase in the driving
required before depinning takes place.

Our results show that the bifurcation from 2d ridges to 3d drop-like
states is subcritical, at least for the parameter values we use. As a
result the 3d states are initially unstable, although with increasing
amplitude the 3d states turn around towards larger values of $\mu$ and
may acquire stability once they resemble steady rivulets. However,
once the 2d ridges near the 2d sniper are unstable to 3d states,
the 2d depinning mechanism no longer leads to stable stick-slip motion
of the ridges, and instead we find a stable 3d version, which we call
stick-slip drops. Our numerics suggest that in both the 2d and 3d
cases the SSR and SSD states terminate in global bifurcations
involving equilibria of saddle-type. However, the theory of such
bifurcations shows that the periodic orbits involved in the global
bifurcations are unstable and indeed our computations in the 2d case
provide some evidence of a saddle-node bifurcation in the neighborhood
of the putative global bifurcation at which the stable SSR states lose
stability. Unfortunately we are unable to follow unstable
time-periodic states such as the SSR and SSD states to confirm the
presence of such global bifurcations.

We were able, however, to interpret the transitions we observed using
a two-dimensional model system analyzed by Krauskopf and Oldeman
\cite{KrOl04}.  This system models the dynamics arising from the
interaction of a sniper and a pitchfork, and its analysis reveals the
great wealth of behavior available to systems of this type. We expect
that transitions present for other values of our parameters can
likewise be interpreted in terms of this model system.  Model systems
such as Eqs.~(\ref{nf1})-(\ref{nf2}) suffer from a drawback, however,
in that the dynamics is necessarily two-dimensional and hence no chaos
is present in the model. A three-dimensional extension of the model
\cite{KrOl06} does exhibit chaotic dynamics associated with the
plethora of global bifurcations revealed in \cite{KrOl04} and offers a
glimpse into the potential behavior associated with the (formally
infinite-dimensional) depinning problem.  However, our simulations
have provided no conclusive evidence for chaotic dynamics. In some
regimes we have also found a Hopf bifurcation in the 2d depinning
problem, usually very close to both the saddle-node of the ridge
states and to the 3d instability, suggesting that the full dynamics of
the system can be captured by examining the interaction of the
Takens-Bogdanov bifurcation with the pitchfork leading to 3d states
\cite{LaZh99}, with reinjection as in \cite{KrOl04}.
 
In the parameter regime where the long-wave approximation applies
the problem studied here, viz. film flow and drop motion on a 
heterogeneous substrate with a well-defined spatial period of the 
wettability, is closely related to the corresponding flow on the
outside or inside of a rotating horizontal cylinder \cite{Thie10b}. In particular,
the depinning dynamics of a two-dimensional drop of partially wetting
liquid on the surface of a rotating cylinder has a close counterpart 
in the depinning dynamics via a sniper bifurcation described here for
two-dimensional drops on heterogeneous substrates \cite{ThKn06,Thie10b}.  
Further exploration of the analogy between these two systems may
therefore be fruitful. In particular, it may be possible to relate the
three-dimensional structures and transitions described in the
present paper to flows on rotating cylinders. For instance, the
transition from stick-slip drops to spatially modulated rivulets via
a Hopf bifurcation (Fig.~\ref{fig:cpnel32}) would correspond to a
transition from drops rotating with the cylinder (a state that has
apparently not been described in the literature on film flow on the
outside of a cylinder) to azimuthal rings \cite{Moff77}.
  
Experiments on pattern formation in flows of thick films on the 
inner surface of a rotating cylinder report a variety of different 
structures and transitions that resemble some of our results
\cite{Melo93,ThMa97}. These include, for instance, stationary straight 
and wavy fronts (similar to our steady spanwise invariant ridge 
and spanwise modulated ridge states).  The latter are called 
``shark teeth'' in \cite{ThMa97}.  At higher rotation velocities the 
stationary wavy fronts decay into drops that are ``dragged up the 
receding wall'' \cite{Melo93}, a transition corresponding to what we 
describe here as a 3d depinning transition. Stationary localized
bumps, i.e., pinned drops, may also be present~\cite{Melo93}.

\acknowledgments

We acknowledge support by the European Union via the FP7 Marie Curie
scheme [Grant PITN-GA-2008-214919 (MULTIFLOW)], the Deutsche
Forschungsgemeinschaft under grant SFB 486, project B13, and by the
National Science Foundation under grant DMS-0908102.


\end{document}